\documentclass[11pt,onecolumn]{IEEEtran}

\usepackage [cmex10]{amsmath}

\usepackage{epsfig,psfig}
\usepackage{graphicx}
\usepackage{amssymb}
\usepackage{graphicx}
\usepackage[english]{babel}

\begin{document}

\title{CMB map restoration}

\author{ J.Bobin \IEEEauthorrefmark{1} \IEEEauthorrefmark{2}, J.-L. Starck \IEEEauthorrefmark{2}, F. Sureau \IEEEauthorrefmark{2} and , J. Fadili \IEEEauthorrefmark{3}
\thanks{ \IEEEauthorrefmark{2} Laboratoire AIM,
IRFU, SEDI-SAP, Service d'Astrophysique,
Orme des merisiers, Bat 709, piece 282
91191 Gif-sur-Yvette, France. \IEEEauthorrefmark{3} GREYC
CNRS-ENSICAEN-Universit de Caen
6, Bd du Marchal Juin,
14050 Caen Cedex
France. \IEEEauthorrefmark{1} Corresponding author : jbobin@cea.fr }}

\maketitle


 \begin{abstract} 
  Estimating the cosmological microwave background is of utmost importance for cosmology. However, its estimation from full-sky surveys such as WMAP or more recently Planck is challenging~: CMB maps are generally estimated via the application of some source separation techniques which never prevent the final map from being contaminated with noise and foreground residuals. These spurious contaminations whether noise or foreground residuals are well-known to be a plague for most cosmologically relevant tests or evaluations; this includes CMB lensing reconstruction or non-Gaussian signatures search. Noise reduction is generally performed by applying a simple Wiener filter in spherical harmonics; however this does not account for the non-stationarity of the noise. Foreground contamination is usually tackled by masking the most intense residuals detected in the map, which makes CMB evaluation harder to perform. In this paper, we introduce a novel noise reduction framework coined LIW-Filtering for Linear Iterative Wavelet Filtering which is able to account for the noise spatial variability thanks to a wavelet-based modeling while keeping the highly desired linearity of the Wiener filter. We further show that the same filtering technique can effectively perform foreground contamination reduction thus providing a globally cleaner CMB map. Numerical results on simulated but realistic Planck data are provided. 
 \end{abstract}

\section{Introduction}
\label{sec:intro}

In mid 2009, ESA (European Spatial Agency) put in orbit the latest space observatory Planck to investigate the  Cosmological Microwave Background (CMB). These data are of particular scientific importance as it will provide more insight into the understanding of the birth of our Universe. Most cosmological parameters can be derived from the study of these CMB data. After a series of  successful CMB experiments (Archeops, Boomerang, Maxima, COBE and WMAP \cite{WMAP}), the Planck ESA mission is providing more accurate data which will be the reference full-sky high resolution CMB data for the next decades. More precisely, recovering useful scientific information requires disentangling the CMB from the contribution of several astrophysical components namely Galactic emissions from dust and synchrotron, Sunyaev-Zel'dovich (SZ) clusters, Free-Free emissions, CO emission to only name a few, see \cite{Bouchet}. Classically, the CMB map is estimated via the application of some source separation methods. However, the application of very sophisticated source separation methods (see \cite{Leach08,NeedletILC,bobin:gmca_itip,BobinICIP11}) does not prevent the final estimated map from being contaminated with various spurious components~: i) instrumental noise and ii) foreground residuals. Dealing with these various sources of contaminations is of paramount importance for most cosmologically relevant tests or evaluations~: this includes CMB lensing reconstruction \cite{CMBLens} or non-Gaussian signatures search. Tackling the problem of noise and foreground residual reduction is therefore important and is at the heart of this paper.\\

\paragraph{Instrumental noise} 

\begin{figure}[htb]
\centerline{\includegraphics[scale=0.5]{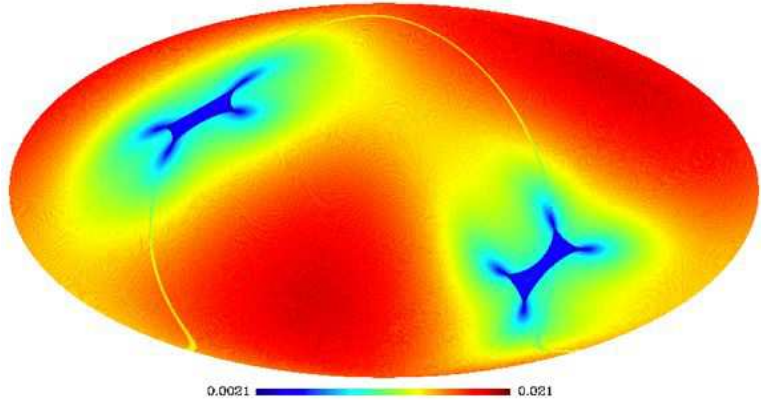}}
\caption{{\bf Instrumental noise in Planck} - simulated noise root mean square map at $217$GHz in mK (Antenna temperature).}
\label{fig:hitmap}
\end{figure}

The very specific scanning pattern in full-sky CMB experiments leads to an instrumental noise closely Gaussian that exhibits significant spatial variation~: Planck noise is far from being stationary. As an illustration, Figure~\ref{fig:hitmap} displays the root mean square (RMS) map of Planck expected instrumental noise at $217$GHz. Furthermore,  while noise is insignificant at large scale, it is largely predominant at small scale (\textit{e.g.} typically for $\ell > 1500$ for Planck).\\
Most source separation methods applied so far to the Planck data are linear methods~: the estimated CMB map can be expressed as a linear combination of all the pixels of the input observations. As such, the propagated instrumental noise in the CMB maps is also typically assumed non-stationary Gaussian.\\
Noise can be a significant limitation for most cosmological tests or reconstructions. High noise level can hamper non-gaussianity tests as the noisy map is likely to be ``more Gaussian" than the noise-free map. The classical approach in the field of cosmology generally consists in reducing the noise of the CMB via a Wiener filter in spherical harmonics. If $C_\ell$ stands for the theoretical power spectrum of the CMB and $C^n_\ell$ the power spectrum of the noise, the Wiener filter is defined as follows~:
\begin{equation}
\forall \ell,m > 0; \quad a^{\hat{x}}_{\ell m} = \frac{C_\ell}{C_\ell + C^n_\ell} \, a^y_{\ell m}
\end{equation}
where $\{a^{\hat{x}}_{\ell m} \} _{\ell,m}$ (\textit{resp.} $\{a^{y}_{\ell m} \} _{\ell,m}$) stands for the spherical harmonics coefficients of the filtered map $\hat{x}$ (\textit{resp.} the input map $y$). While this approach makes profit of some knowledge of the energy distribution of the noise in frequency, it does not account for its spatial variations or non-stationarity. It is very likely that such a non-stationarity may create global non-gaussian signatures at high frequency when such variations are sharp. Strong non-stationarities are known to have a strong impact on CMB lensing reconstruction (see \cite{LensingIn09}), creating an undesired reconstruction bias.\\
Another less traditional but commonly encountered noise reducing technique is the local Wiener filtering. It boils down to apply the Wiener filter in the Fourier space on patches in the pixel domain. The main drawback of this method is that while large patches should be favored to capture the non-stationarity of noise, they are enable to capture correlations at scales larger than the patch size.\\

\paragraph{Foreground residual}  
The first evaluation of already sophisticated CMB-dedicated source separation methods has been performed in \cite{Leach08}. Since, novel very effective methods have been proposed Needlet ILC \cite{NeedletILC}, Generalized ILC \cite{GenILC}, SMICA \cite{ica:Del2003} or GMCA \cite{BobinICIP11}. However, due to the very high complexity of the data, none of these very effective methods is able to provide a CMB map that is guaranteed to be free of foreground contaminations. If the residual contamination is generally low at large scales ( {\it e.g.} for $\ell < 500$ for Planck), it is generally significant at higher frequencies. It is likely that the remaining contamination mainly originates from point sources or strong features in the galactic plane. Obviously the presence of these residuals highly biases any cosmological tests to be performed on contaminated CMB map. The usual approach consists in masking the known spurious pixels of the CMB prior to any post-processing (\textit{e.g.} $20$ to $40\%$ of the sky are usually masked). However masking raises several issues~: i) it limits the number of samples to which any test can be applied, limiting its statistical relevance and ii) masking generally makes any post-processing much more complicated to perform properly (\textit{e.g.} inpainting of the mask is generally needed prior to any CMB lensing reconstruction \cite{CMBLens}). Avoiding or at least limiting the extent of the mask to be applied should greatly help improving any cosmological test to be made on the estimated CMB map. Furthermore, masking does not prevent from the presence of residuals of undetected point sources which are likely be be numerous all over the sky. Further reducing the amount of foreground residuals should clearly be helpful.\\

\paragraph{Contribution of this paper}
The contribution of this paper is twofold~: 
\begin{itemize}
\item It introduces a novel noise reduction framework that, in opposition to the classical spherical-harmonics based Wiener filter, accounts for the non-stationarity of the noise. This new method preserves the linearity of the filtering. In fact, linearity is crucial in this field as it makes the study of error propagation via Monte-Carlo simulations much more convenient to carry out. In this framework, we make use of a simple, but efficient, wavelet-based modeling of the noise that allows for the modeling of non-stationarities at different scales. This will be discussed in Section~\ref{sec:varestim}
\item We discuss an extension of this method to further estimate an approximate contribution of the foreground residuals per wavelet scales by means of RMS maps. Thereby contamination reduction could be tackled within the same filtering framework.
\end{itemize}
The noise/contamination reduction problem, recast as a quadratically regularized least square problem, is then effectively solved using recently introduced algorithms based on proximal calculus. Extensive numerical experiments are provided that show the effectiveness of the proposed method in reducing noise in Section~\ref{sec:noisereduc} and foreground residuals in Section~\ref{sec:contreduc}.

%
%

\section{CMB map filtering with non-stationary noise}
\label{sec:filtering}

\subsection{CMB map filtering}

In a very general context and more precisely in the context of Planck, it is customary to assume that the input noise data, denotes by $y$ in the following, is modeled as the linear combination of the sought after noise-free signal $x$ and the noise term $n$~:
$$
y = x + n
$$ 
Decorrelation between $x$ and $n$ is also classically assumed. In what follows, we will assume that each of these signals is sampled on the sphere according to the Healpix \cite{Healpix} sampling grid. The sampled CMB signal $x$ is assumed to be composed of $N$ samples $\{x[k]\}_{k =1,\cdots,N}$. Recovering an estimate $\hat{x}$ of $x$ can then be formulated as an inverse problem. According to the Bayesian way of solving inverse problems, some \textit{a priori} knowledge about the signal of interest $x$ has to be expressed. This inference framework is particularly well suited to the CMB denoising issue as strong \textit{prior} assumptions can be formulated. Indeed, since \cite{Komatsu09}, it is well established that the cosmological microwave background can be approximated as an isotropic Gaussian random field with a high accuracy. This particularly means that it is fully characterized by its power spectrum $C_\ell$.\\
More precisely, the CMB $x$ can be defined by its expansion in spherical harmonics~:
\begin{equation}
x = \sum_{\ell = 0}^\infty \sum_{m = -\ell}^\ell a_{\ell m} Y_{\ell m}
\end{equation}
Assuming that the CMB is fully Gaussian with power spectrum $C_\ell$, its spherical harmonics $\{a_{\ell m}\}_{\ell m}$  are independently distributed following a Gaussian distribution with mean $0$ and variance $C_\ell$~:
\begin{equation}
a_{\ell m} \sim \mathcal{N}\left( 0 , C_\ell \right)
\end{equation}
The power spectrum of the CMB is generally estimated with the pseudo power spectrum defined as follows~:
\begin{equation}
\hat{C}_\ell = \frac{1}{2\ell + 1}\sum_{m = -\ell}^\ell a_{\ell m}^\star a_{\ell m}
\end{equation}
where $a_{\ell m}^\star$ stands for the conjugate of $a_{\ell m}$. The \textit{a priori} probability distribution $\mathcal{P}$ of the CMB map $x$ is a normal distribution with covariance matrix $\mathcal{F}^H {\bf C} \mathcal{F} $ where $\mathcal{F}$ stands for the spherical harmonics basis and $\mathcal{F}^H$ its adjoint. The matrix $\bf C$ is diagonal with ${\bf C} = \mbox{diag}\left( C_\ell \right)$. This yields the following CMB \textit{prior} probability distribution~:
\begin{equation}
\mathcal{P}(x) \propto \exp - x^T \mathcal{F}^H {\bf C}^{-1} \mathcal{F} x 
\end{equation}
In this section, we assume that the instrumental noise can be modeled as a non-stationary but uncorrelated Gaussian field in the pixel domain such that~:
$$
\forall k=1,\cdots,N; \quad n[k] \sim \mathcal{N}\left(0,\sigma_k^2\right)
$$
Written differently, $n$ is a realization of a multivalued Gaussian variable with mean $0$ and covariance matrix ${\bf \Sigma} = \mbox{diag}\left( \sigma_1^2 \cdots \sigma_N^2 \right)$. The case of correlated noise will be discussed later on in this section. From the Gaussian modeling of $n$, the likelihood function is trivially obtained as follows~:
\begin{equation}
\mathcal{L}(y | x ,  {\bf \Sigma}) \propto \exp -\left( y - x \right)^T {\bf \Sigma}^{-1} \left( y - x \right)
\end{equation}
Following Bayes rule, the posterior distribution for $x$ is the following~:
\begin{eqnarray}
P(x | y) & \propto & \mathcal{L}(y | x , {\bf \Sigma}) \mathcal{P}(x) \\ 
&\propto& \exp - \left [  x^T \mathcal{F}^H {\bf C}^{-1} \mathcal{F} x +  \left( y - x \right)^T {\bf \Sigma}^{-1} \left( y - x \right) \right ]
\end{eqnarray}
The CMB map can be estimated using the maximum a posteriori (MAP) estimator with the following analytical solution~:
\begin{eqnarray}
\label{eq:map}
\hat{x} & = & \mbox{Argmax}_x \, P(x|y) \\
 & = & \left( {\bf \Sigma }^{-1} +  \mathcal{F}^H {\bf C}^{-1} \mathcal{F} \right)^{-1} {\bf \Sigma}^{-1} y 
 \label{eq:wiener}
\end{eqnarray}
The reader would here recognize a classical Wiener filter. Contrary to the classically used Wiener filter in spherical harmonics, the solution of Equation~\eqref{eq:map} explicitly account for the non-stationarity of the noise. However, the computational evaluation of this Wiener-like solution raises some numerical issues~: the noise covariance matrix is diagonal in the pixel domain while the CMB covariance matrix is diagonal in the spherical harmonics domain. The solution filter that appears in Equation~\eqref{eq:map} is formulated in the pixel domain~:
$$
\left( {\bf \Sigma }^{-1} +  \mathcal{F}^H {\bf C}^{-1} \mathcal{F} \right)^{-1} {\bf \Sigma}^{-1} 
$$
In this expression, the matrix $\mathcal{F}^H {\bf C}^{-1} \mathcal{F}$ is the inverse covariance matrix of the CMB in the pixel domain. This matrix is clearly not diagonal. Recalling that $N^2$ is of the order of  $1e13$ for WMAP and $2e15$ for Planck, storing and handling this matrix is way too unrealistic. As an alternative, the most effective approach in this situation would amount to solve the linear system of equations $\left( {\bf \Sigma }^{-1} +  \mathcal{F}^H {\bf C}^{-1} \mathcal{F} \right) \hat{x} =  {\bf \Sigma}^{-1} y$ using a conjugate gradient (CG - \cite{miki:Golub}) for instance. However, in the next section, we will rather make use of recently introduced proximal algorithms. This optimization framework leads to algorithms which exhibits a lower speed of convergence than CG but with the gain of more flexibility to handle different filtering techniques or better constrain the solution. \\

\subsubsection{Iterative filtering}
As said clearly in the previous paragraph, computing the MAP solution requires inverting a system of equations which turns to be intractable due to the large scale of Planck data. The most straightforward numerical solver is the well-known conjugate gradient. However, while this solver is known to be a fast and accurate numerical method for solving linear systems of equations, it lacks the flexibility of more modern optimization techniques. For this reason we rather opt for the more flexible optimization framework of proximal calculus (see \cite{CombettesWajs05} and references therein).\\

\paragraph{A flexible optimization framework - forward backward splitting}
The problem in Equation~\eqref{eq:wiener} can be written as minimization of  the sum of two operators $f_1$ and $f_2$~:
\begin{eqnarray}
\label{eq:prox}
\hat{x} & = & \mbox{Argmin}_x \, f_1(x) + f_2(x) \\
& = &   x^T  \mathcal{F}^H {\bf C}^{-1} \mathcal{F} x + \left( y - x \right)^T {\bf \Sigma}^{-1} \left( y - x \right)
\end{eqnarray}
The function $f_1(x) = x^T  \mathcal{F}^H {\bf C}^{-1} \mathcal{F} x$ is convex and lower semi-continuous and differentiable; the function $f_2(x) = \left( y - x \right)^T {\Sigma}^{-1} \left( y - x \right)$ is strictly convex, continuous and differentiable. Such problem as therefore a unique solution (see  \cite{CombettesWajs05} ).\\
If $\nabla f_2(x)$ denotes the gradient of $f_2$, we will also assumed that this gradient is $L$-Lipschitz. In the specific setting of interest in this article, $\nabla f_2(x) = - 2  {\Sigma}^{-1} \left( y - x \right)$ and $L = 2\| \Sigma ^{-1}\|_2$. In this context, an effective algorithmic framework is the forward-backward splitting (FPS) technique defined in \cite{CombettesWajs05}. In the paradigm of proximal calculus, any convex lower semi-continuous  function (see \cite{CombettesWajs05}) admits a well defined proximal operator~:
\begin{equation}
\label{eq:pop}
\mbox{prox}_{\gamma f}(z) = \mbox{Argmin}_u \gamma f(u) + \frac{1}{2} \|z - u \|_{\ell_2}^2
\end{equation}
In the FBS framework, the solution to the problem in Equation~\eqref{eq:prox} is computed via an iterative fixed point algorithm such that at iteration $k$~:
\begin{eqnarray}
\label{eq:FBS}
x^{k+1} & = & \mbox{prox}_{\gamma f_1}\left(x^k - \gamma \nabla f_2(x^k) \right) \\
& = & \mbox{prox}_{\gamma f_1}\left(x^k + 2 \gamma {\Sigma}^{-1} \left( y - x^k \right)\right) 
\end{eqnarray}
As proved in \cite{CombettesWajs05}, the convergence of the above fixed point algorithm is guaranteed under the condition $\gamma < \frac{2}{L}$; ; note that in the setting of CMB map recovery with uncorrelated non-stationary noise, $L = 1/\min_k\sigma_k^2$.\\

\paragraph{The proximal operator of $f_1$} In the problem of interest in this paper, $f_1(x) =  x^T  \mathcal{F}^H {\bf C}^{-1} \mathcal{F} x$. As this function is continuous and differentiable, it is \textit{a fortiori} lower semi-continuous. Following the definition in Equation~\eqref{eq:pop}, the proximal operator of $f_1$ is defined as follows~:
\begin{equation}
\mbox{prox}_{\gamma f_1}(z) = \mbox{Argmin}_u \gamma u^T  \mathcal{F}^H {\bf C}^{-1} \mathcal{F} u + \frac{1}{2} \|z - u \|_{\ell_2}^2
\end{equation}
The solution to this problem has a simple closed-form expression~:
\begin{eqnarray}
\mbox{prox}_{\gamma x^T  \mathcal{F}^H {\bf C}^{-1} \mathcal{F} x} & = &  \left( I+  2 \gamma \mathcal{F}^H {\bf C}^{-1} \mathcal{F} \right)^{-1} z\\
& = & \mathcal{F}^H \left( I+  2 \gamma {\bf C}^{-1}  \right)^{-1} \mathcal{F} z
\label{eq:proxw}
\end{eqnarray}
The proximal operator of $f_1$ is a mere Wiener filter in the spherical harmonics space.\\

To sum everything up, the resulting \textit{iterative} Wiener-like filtering is as follows~: \\

\begin{center}
\centering
\begin{tabular}{|c|} \hline
\label{algo:wiener}
\begin{minipage}[hbt]{0.95\linewidth}
\vspace{0.15in}
\footnotesize{
\textsf{{\bf 1. Initialization~:} Set the number of iterations $I_{\max}$, the step size $\gamma = \min_k \sigma_k^2$ and the starting point $x^{0}$}\\

\textsf{{\bf 2. Main loop~:} apply iteratively the following steps at each iteration $k$~:} \\

\hspace{0.2in} \textsf{a -- Forward step of the FPS~:}\\

\hspace{0.4in}$ u  = x^k + 2 \gamma {\bf \Sigma}^{-1} \left( y - x^k \right)$\\

\hspace{0.2in} \textsf{b -- Backward step of the FPS~:}\\

\hspace{0.4in}$x^{k+1} = \mathcal{F}^H \left( I+  2 \gamma {\bf C}^{-1}  \right)^{-1} \mathcal{F} u $\\

\textsf{{\bf 3. Stop} when $k = I_{\max}$.}}

\vspace{0.15in}
\end{minipage}
\\\hline
\end{tabular}
\vspace{0.1in}
\end{center}

\paragraph{Computational cost} The computational cost of this iterative Wiener algorithm is particularly low. The first step (2 -- a) only requires entrywise multiplications and additions of vectors. It is therefore of the order of $\mathcal{O}(N)$. The second step (2 -- b) is by far the most computationally demanding; it requires entrywise multiplications and additions of vectors of the order of $\mathcal{O}(\ell_{\max})$ (where generally $\ell_{\max} \simeq 3000$) and single forward/backward spherical harmonics transforms. In the numerical experiments of Section~\ref{sec:noisereduc} and \ref{sec:contreduc}, the starting point $x^{(0)}$ will be chosen as having entries equal to zero. The total number of iterations will be no greater than $I_{\max} = 250$.\\

\paragraph{Linearity of the iterative Wiener filter}
The proposed iterative filtering technique preserves the linearity of the overall CMB map processing. In fact, in every step of the proposed algorithm, each output depends linearly on their inputs. This property is essential as it allows for a simple way to study the propagation of errors via Monte-Carlo simulations; each simulation then has to undergo exactly the same processing step described above.

\subsection{The case of correlated noise}

In most CMB surveys, such as WMAP or Planck to only name two, the instruments have been thoroughly calibrated on ground thus meaning that the noise level of these instruments is accurately known. Assuming a perfect calibration of the data, the non-stationarities of instrumental noise mainly originate from the non-uniform scanning of the sky. The lower the number of scanning time in one area is, the higher the noise level is and \textit{vice versa}. Figure~\ref{fig:hitmap} shows the RMS map of noise at channel $217$GHz; mathematically this is no more than the square root of the diagonal of $\bf \Sigma$. However this is also well known that, contrary to what has been assumed in the previous section, the noise is far from being uncorrelated. There are mainly two origins for the correlation of noise~:\\
\begin{itemize}
\item The way the sky is scanned by Planck makes the noise correlated along the scanning direction. This means that the noise is correlated along elongated patterns in the pixel domain; for numerical reasons, this can hardly be accounted for in the pixelwise covariance matrix ${\bf \Sigma}$.\\
\item Most modern CMB-dedicated source separation methods Needlet ILC \cite{NeedletILC}, Generalized ILC \cite{GenILC}, SMICA \cite{ica:Del2003} or GMCA \cite{BobinICIP11} rely on a local analysis of the data in the wavelet or needlet domain. Furthermore, the final CMB map is generally obtained as the mixture of observations that do not share the same resolution. This means that the noise that finally contaminates the CMB is, in the pixel domain, correlated.\\
\end{itemize}
It then seems natural to model non-stationary correlated noise~: i) in the wavelet domain to capture its correlation and ii) locally to capture its non-stationarity.\\

\subsubsection{Wavelets - the basics}

Wavelets are well known to be the tool of choice to analyze non-stationary signals \cite{ima:mallat98}. When considering spherical data, there exist several implementations of wavelets on the sphere. In this paper, we will make use of the isotropic undecimated wavelet transform (IUWT) introduced in \cite{starck:sta05_2}. The IUWT can be seen as an extension of the celebrated \textit{\`a trous} algorithm to spherical data (see the seminal book \cite{ima:mallat98} and references therein).\\
As any wavelet transform, the IUWT is first defined by a scaling function $\phi_{\ell_c}(\theta,\varphi)$ defined by a fixed frequency cut-off $\ell_c$. As emphasized in \cite{starck:sta05_2}, this scaling function is built so that it is azimuth-invariant on the sphere (\textit{i.e.} it does not depend on $\varphi$). Classically, each rescaled version of the scaling function $\phi_{\frac{\ell_c}{2^j}}$ is associated to a low-pass filter $h_j$ or equivalently $H_j(\ell)$ in spherical harmonics. Traditionally, the wavelet-based multiscale analysis is performed by computing successive smooth approximations of $x$. These smooth approximations are obtained by convolving $x$ with dyadically rescaled versions of the scaling function~:
$$
\forall j=0,\cdots,J; \quad c_j = h_j \star x
$$
where $J$ is the total number of scales. The so-called wavelet function at scale $j$, $\psi_{\frac{\ell_c}{2^j}}(\theta,\varphi)$, is defined as the difference of two consecutive rescaled scaling functions~:
$$
\psi_{\frac{\ell_c}{2^j}}(\theta,\varphi) = \phi_{\frac{\ell_c}{2^{j-1}}}(\theta,\varphi) - \phi_{\frac{\ell_c}{2^j}}(\theta,\varphi)
$$ 
The wavelet function at scale $j$ can be equivalently associated to a high-pass filter $g_j$ or equivalently $G_j(\ell) = 1 - H_j(\ell)$ in spherical harmonics. The wavelet coefficients at scale $j$ are then computed by convolving the smooth approximation of $x$ at scale $j$ with the $j$-th wavelet function $\psi_{\frac{\ell_c}{2^j}}(\theta,\varphi)$. 
$$
\forall j=0,\cdots,J; \quad w_{j+1} = g_j \star c_j
$$
In spherical harmonics, this amounts to a simple multiplication of the signal's spherical harmonic coefficients with the wavelet filter $G_j$.\\
The decomposition being redundant, several strategies can be designed to reconstruct $x$ from the wavelet coefficients and the coarse resolution $c_J$. The simplest is the one advocated by the \textit{\`a trous} algorithm~:
$$
x = c_J + \sum_{j = 1}^{J} w_j
$$
In the following, noise modeling will be performed in each wavelet scale; \textit{i.e.} on each set of coefficients $\{w_j\}_{j=1,\cdots,J}$.\\

\subsubsection{Wavelet-based statistical modeling of correlated noise}
\label{sec:wtmodel}
As emphasized previously, one elegant way to model the instrumental noise is to consider the distribution of its expansion coefficients in the spherical wavelet domain. The basic idea consists in assuming that the wavelet coefficients of noise are approximately decorrelated. This idea takes its roots in the field of multiscale statistical modeling~: it has been shown that wavelets exhibit (almost)-decorrelating properties for some classes of non-stationary stochastic processes. To give only one example, this is the case for fractional Brownian motion \cite{fBM92} which has been extensively used to model power-law following stochastic processes. In the following, $w^n_{j}[k]$ will denote the wavelet coefficient of the noise term $n$ in scale $j$ at pixel $k$ with the convention~: $j=1$ corresponds to the finest scale and $j=J-1$ to the coarse resolution. From this definition, we will assume that these wavelet coefficients verify~:
\begin{equation}
\mathbb{E}\left \{ w^n_{j}[k] w^n_j[i] \right \} = {\sigma_{j}^n}[k]^2 \delta_{k,i}
\end{equation}
where ${\sigma_{j}^n}[k]^2$ stands for the local variance of the noise at pixel $k$ and $\delta_{k,i}$ for the kronecker delta function. This entails that the covariance matrix of the noise in wavelet scale $j$ is diagonal and equal to~:
\begin{equation}
{\bf \Sigma}_{n,j} = \mbox{diag}\left({\sigma_{j}^n}[k]^2\right)
\end{equation}

The local standard deviation of the noise, ${\bf \Sigma}_{n,j}$, has to be estimated from the data beforehand. In the numerical experiments in Section~\ref{sec:noisereduc}, these parameters are estimated from a single noise realization. Within Planck, it is common to compute the so-called \textit{jack-knife} map defined as the difference between consecutive scanning rings; provided that the instrument calibration do not vary from one ring to the other, it gives a noise realization.\\
 Assuming that the local variances ${\sigma_{j}^n}[k]^2$ of the noise vary slowly from one pixel to another, it can be approximated at pixel $k$ by the empirical estimate of the variance in a fixed surrounding neighborhood; in the following this neighborhood will be a patch of size $\sqrt{P}\times \sqrt{P}$ centered about the pixel $k$~:
\begin{equation}
\label{eq:varest1}
\hat{\sigma}_{j}^n[k]^2 = \frac{1}{P} \sum_{i \in \mathcal{N}[k]} {w^n_{j}}[i]^2
\end{equation}
Where $\mathcal{N}[k]$ stands for the indices of the pixels that compose the patch that surrounds the pixel $k$. Note that the patch size $\sqrt{P}$ will highly depend on how fast the noise variance varies spatially. This particularly means that $\sqrt{P}$ should be smaller in the finest scale and larger when $j$ increases. Following the natural dyadic scaling of the wavelet transform, the patch size in scale $j$ will scale as follows~: $P_j = 4^{j} P_1$ where $P_1$ stands for the patch size at the finest scale.\\

\subsection{Numerical experiments}
\label{sec:noisereduc}

\begin{figure}[htb]
{\includegraphics[scale=0.45]{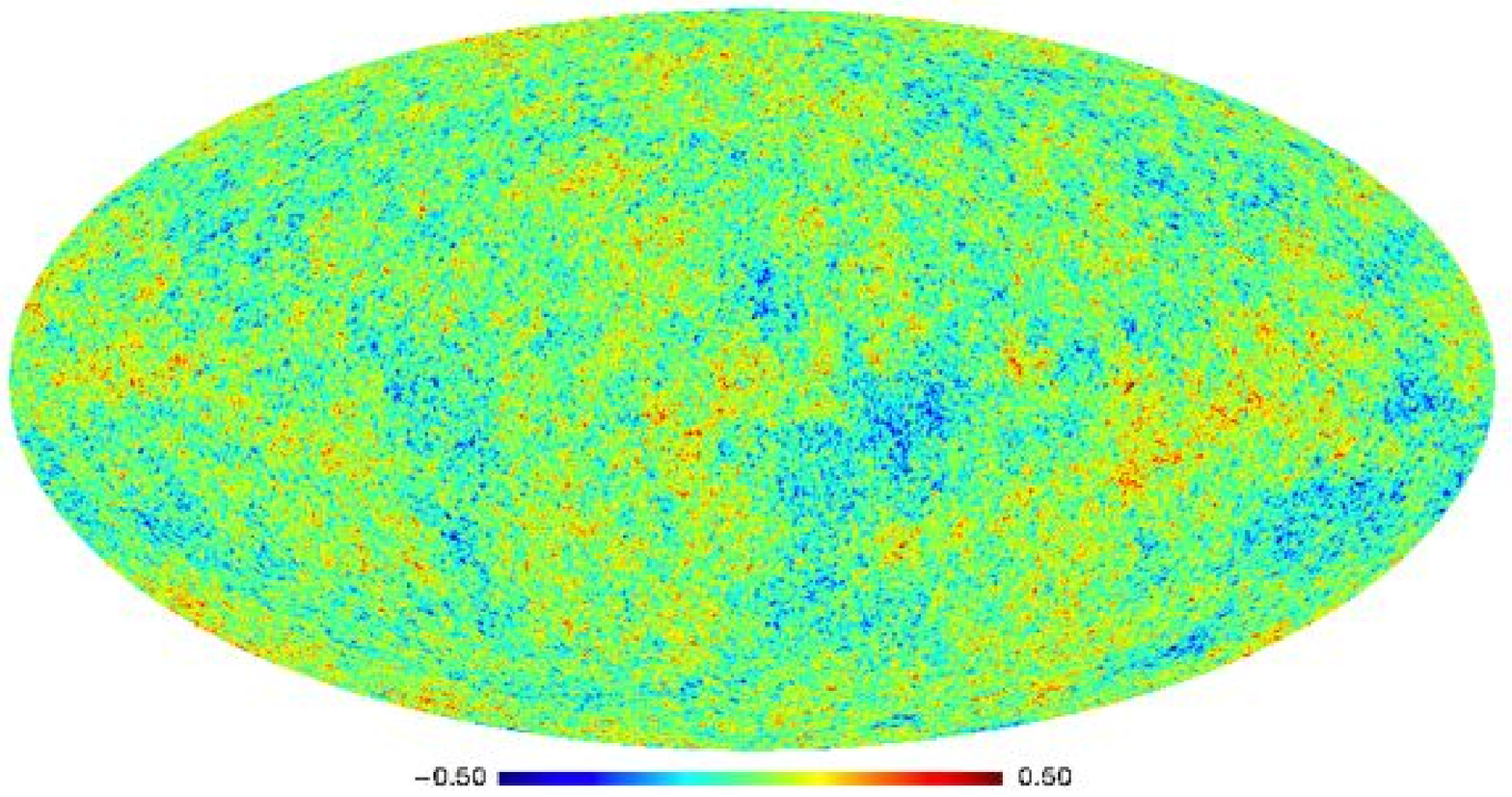}}
\hfill
{\includegraphics[scale=0.45]{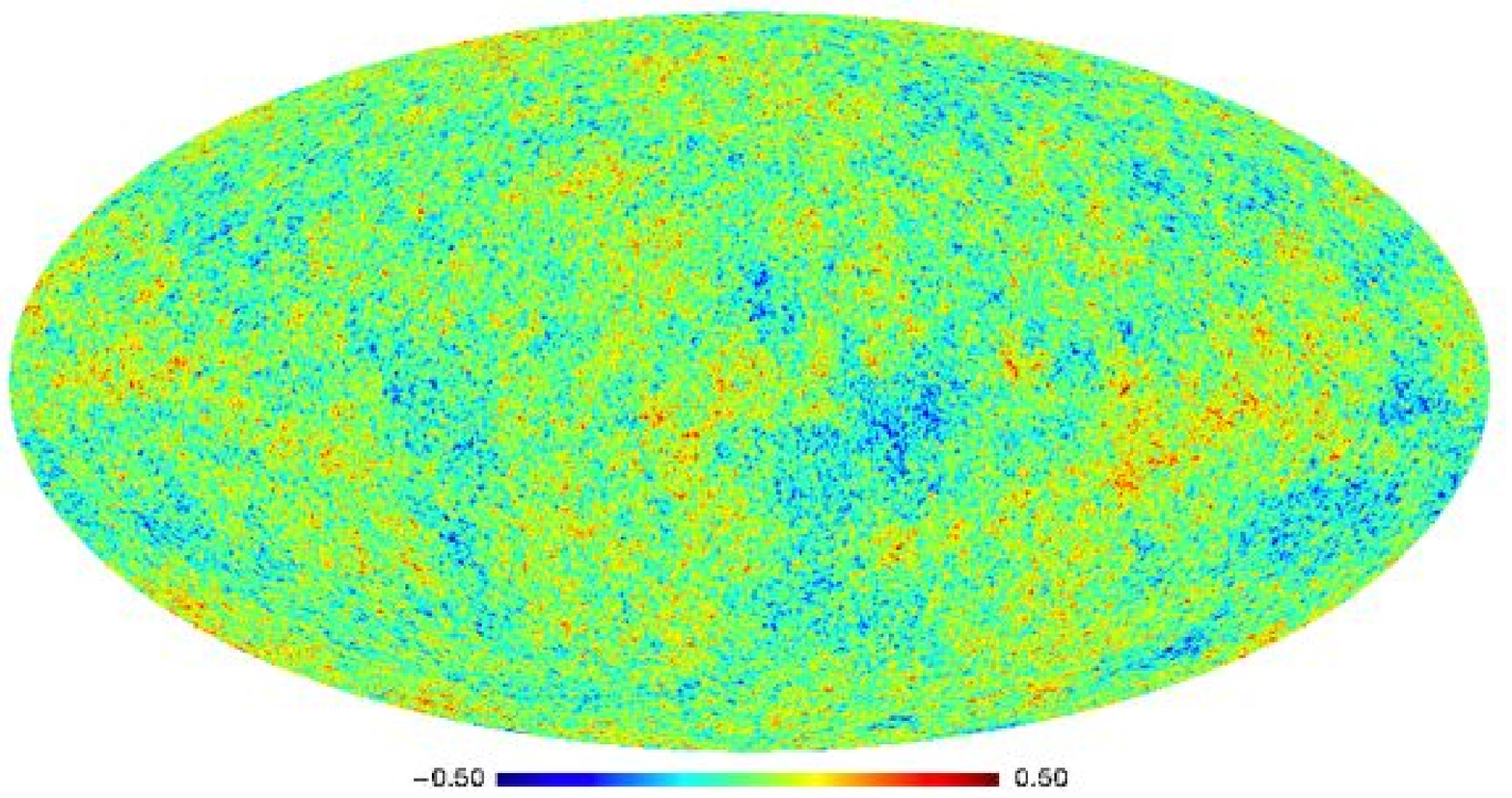}}
\vfill
{\includegraphics[scale=0.45]{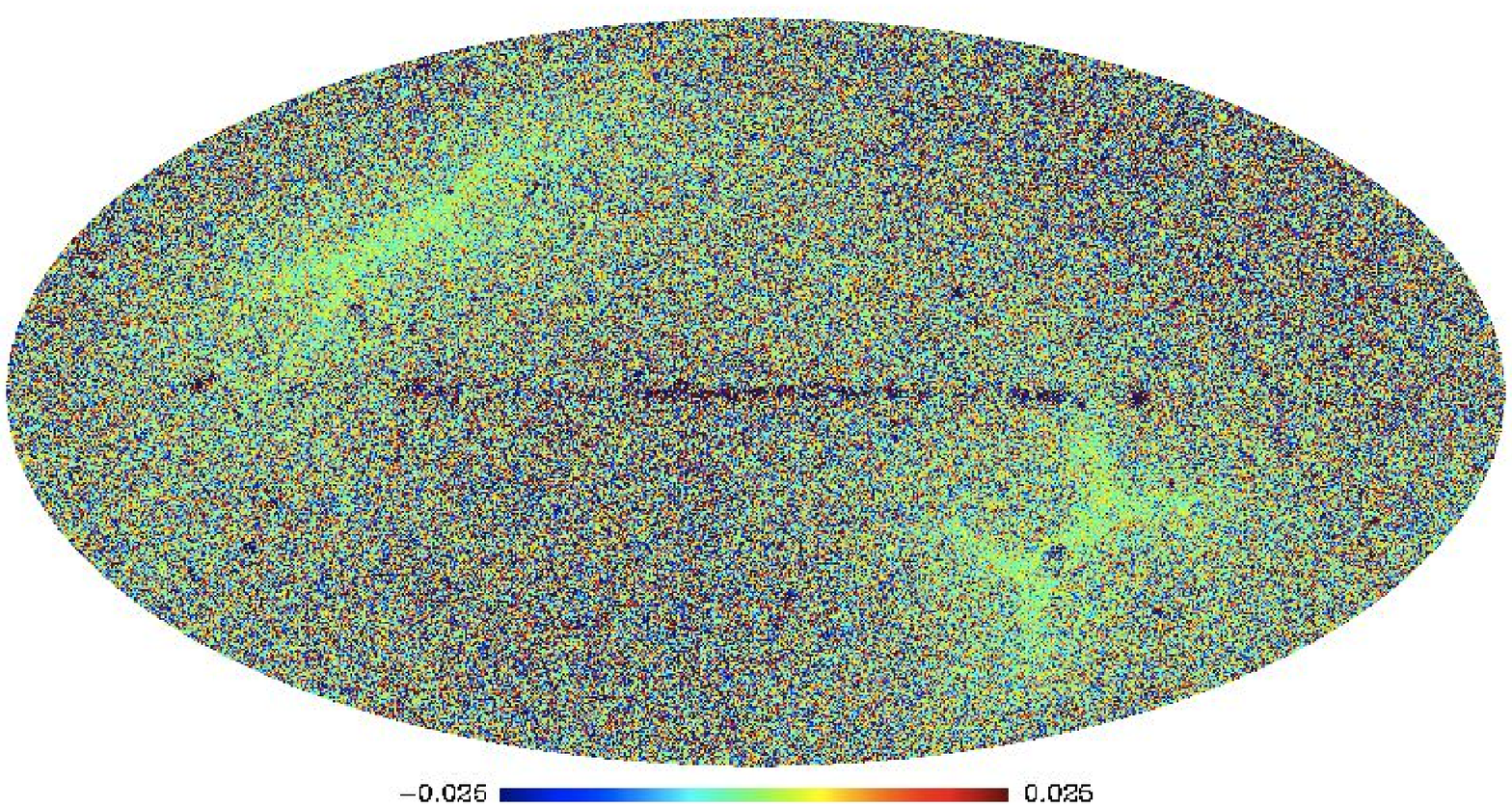}}
\hfill
{\includegraphics[scale=0.45]{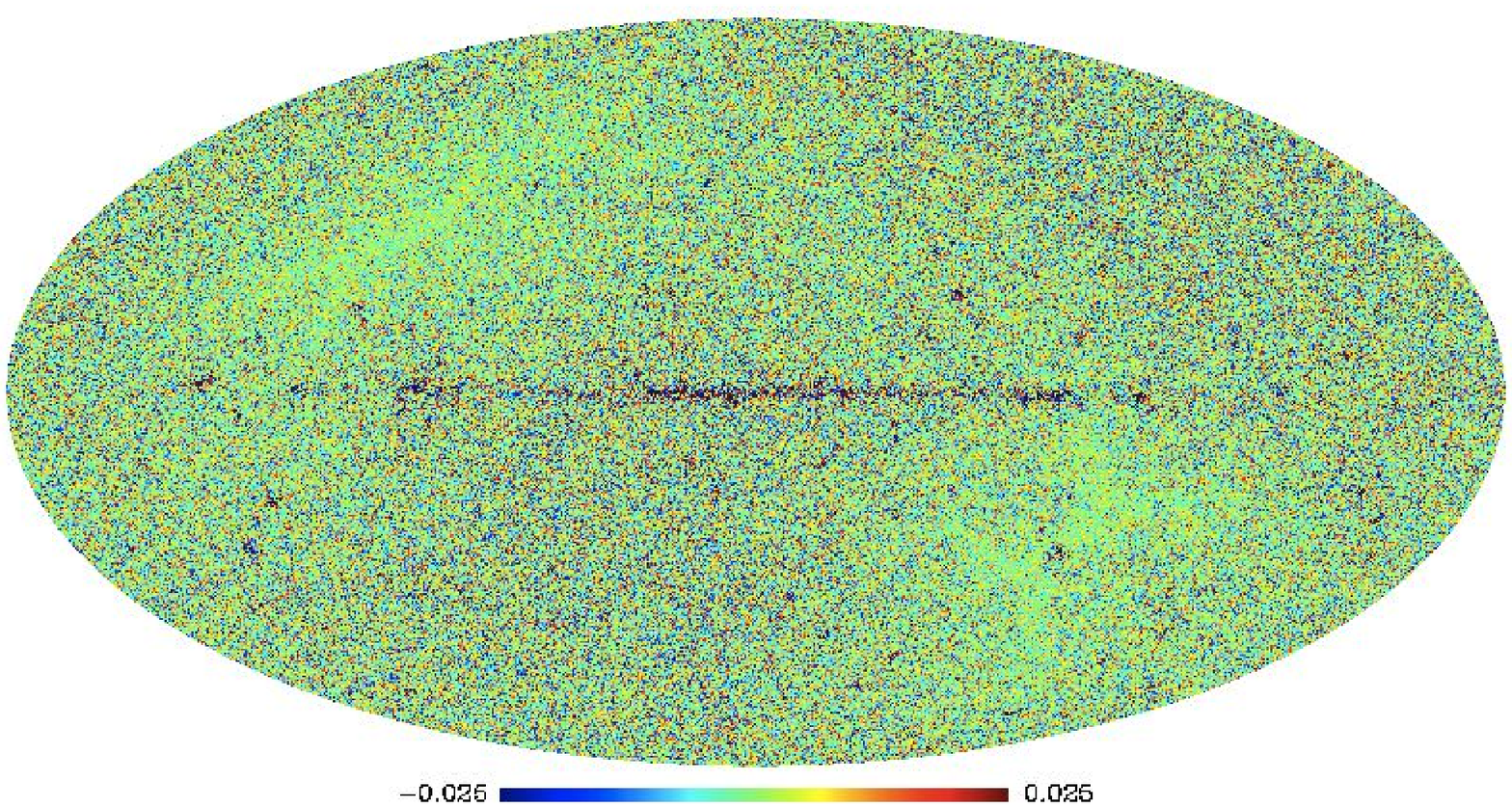}}
\caption{{\bf CMB maps - Top-left~:} Noisy image. {\bf Top-right~:} LIW-Filtering solution. {\bf Bottom-Left~:} Residual map (true map - noisy map) in the finest wavelet scale. {\bf Bottom-Right~:} Residual map (true map - LIW-Filtering solution)  in the finest wavelet scale. Units in $\mbox{mK}$.}
\label{fig:Maps}
\end{figure}

\begin{figure}[htb]
{\includegraphics[scale=0.45]{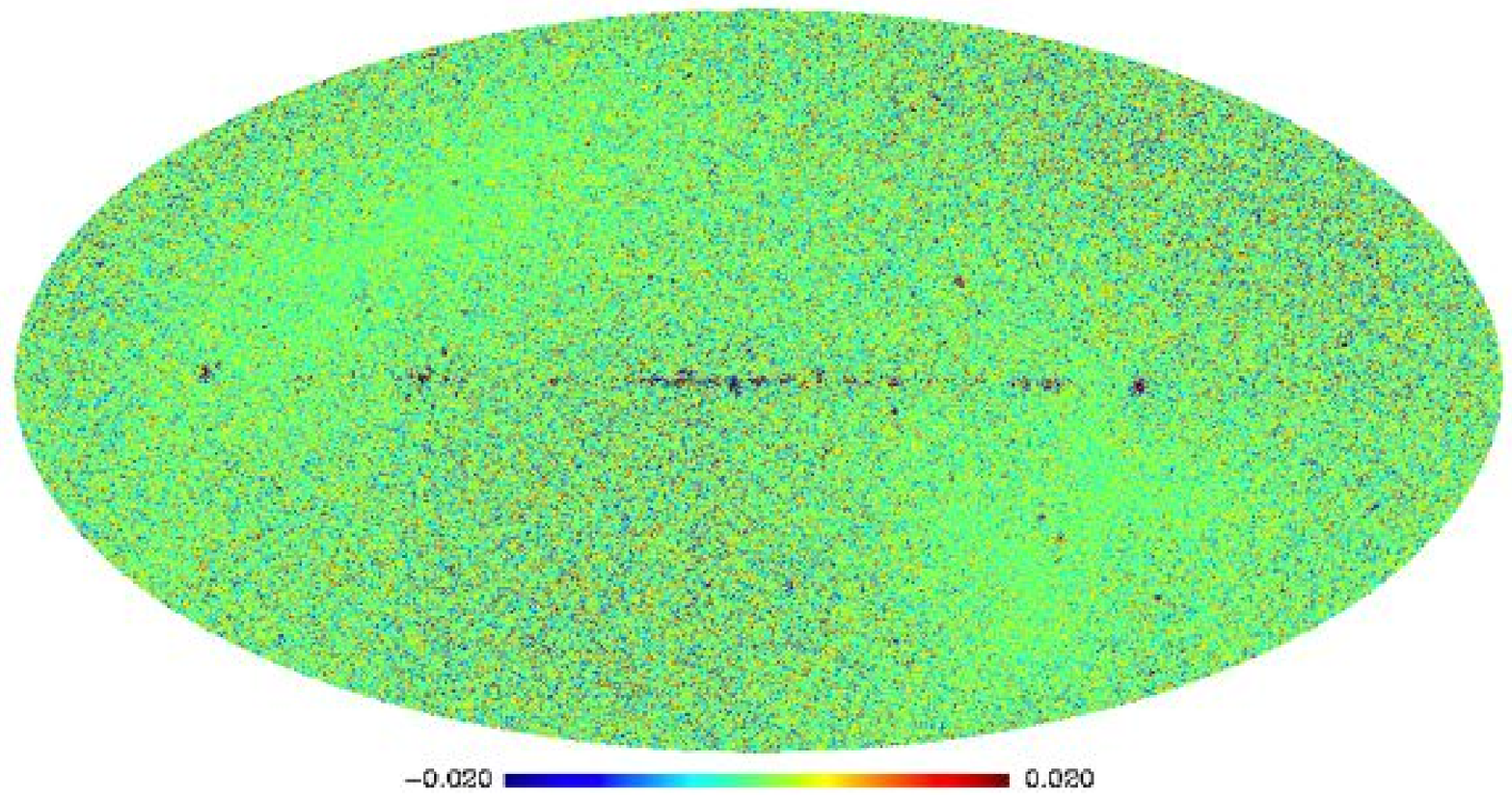}}
\hfill
{\includegraphics[scale=0.45]{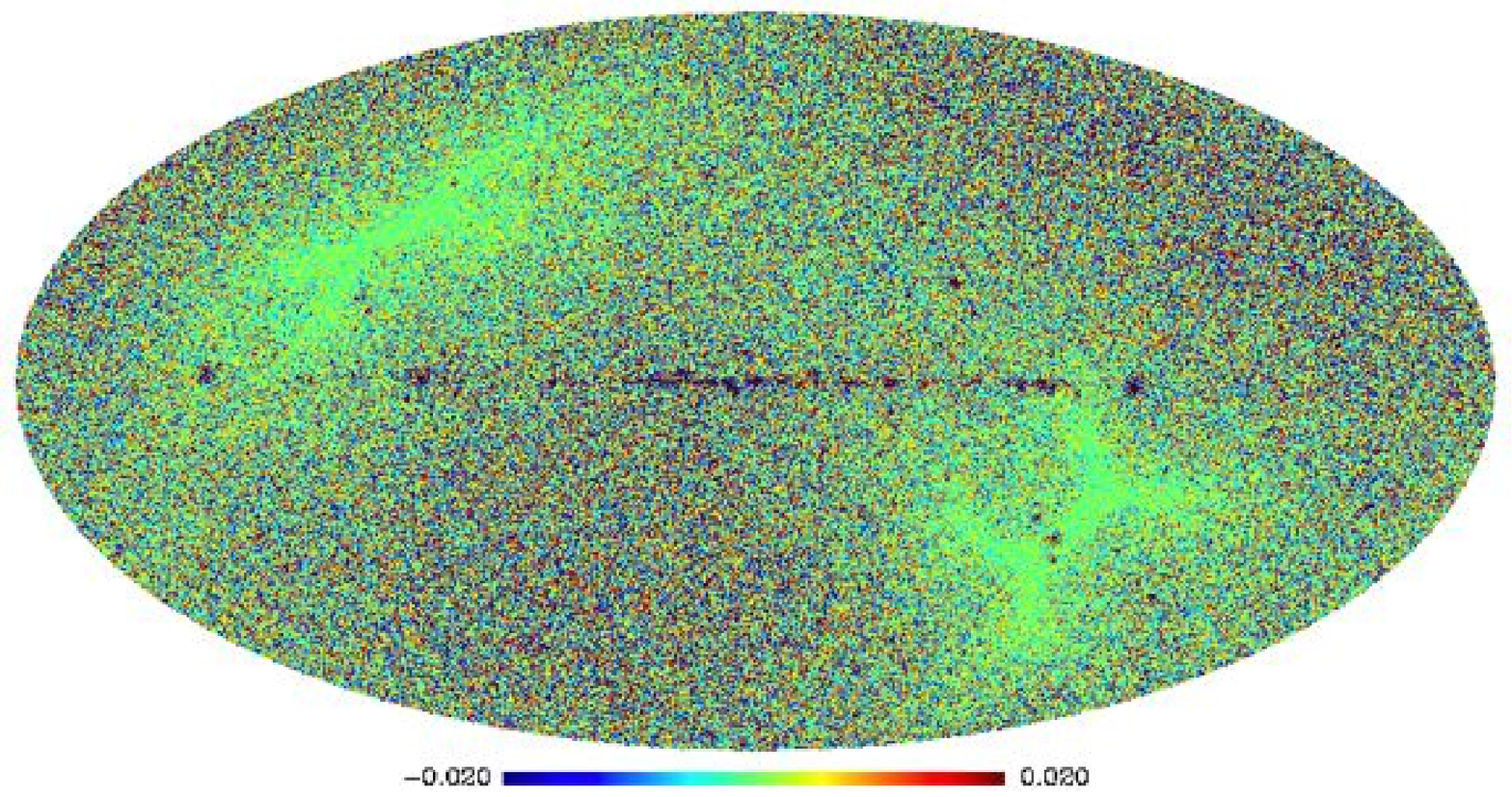}}
\caption{{\bf Removed noise map ({\it i.e.} noisy data - filtered data)  in the finest wavelet scale - Left~:} from the {\it global} Wiener filter. {\bf Right~:} from the LIW-Filtered solution. {\bf These maps correspond to what has been filtered from the noisy map}.  The amount of noise removed by LIW-Filtering is more realistic. Units in $\mbox{mK}$.}
\label{fig:RMaps}
\end{figure}

\begin{figure}[htb]
\centerline{\includegraphics[scale=0.35]{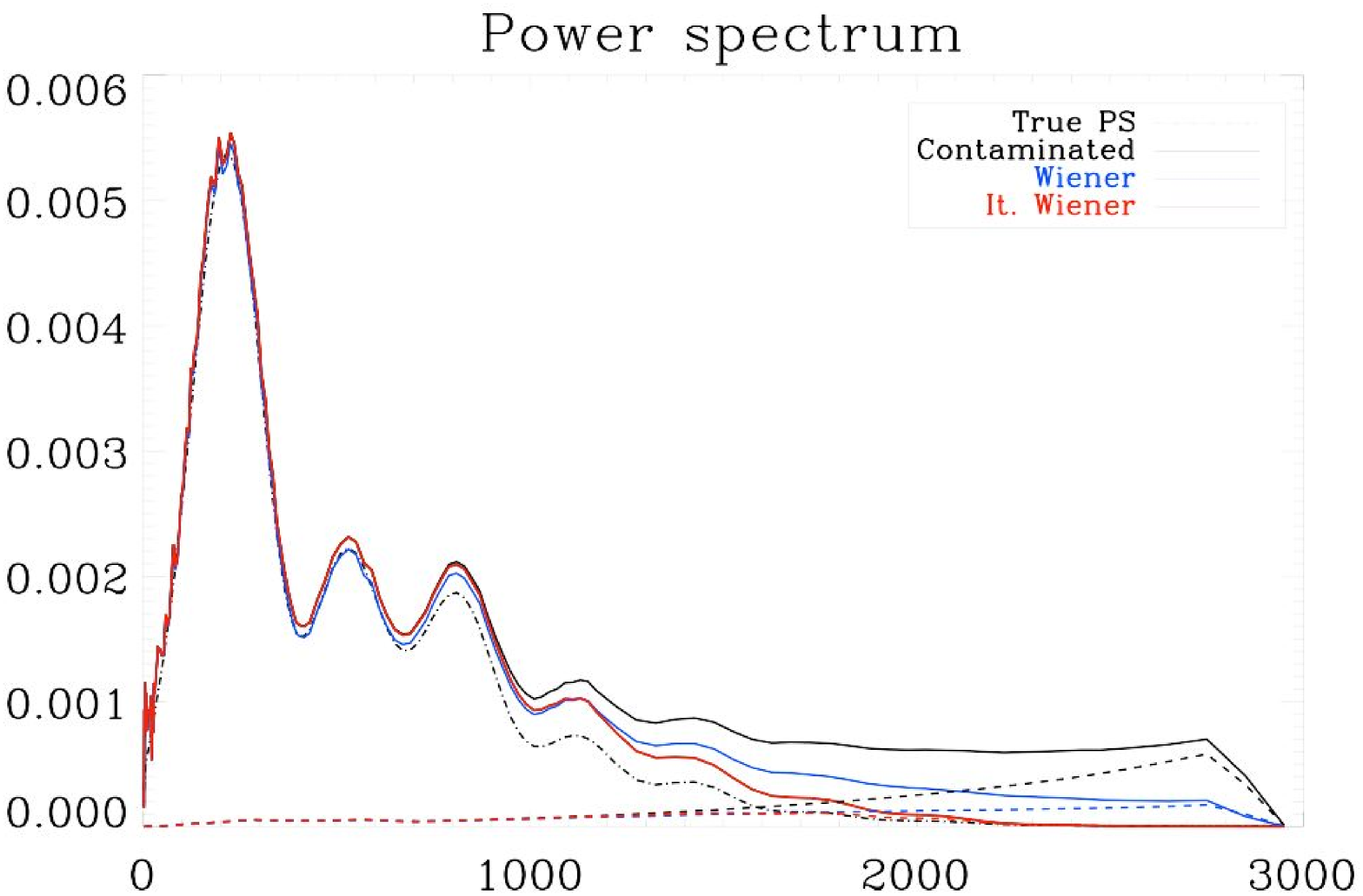}}
\caption{{\bf Power spectra.} In abscissa~: spherical harmonics coefficient. In ordinate~: CMB power in $\mbox{mK}^2$. Dashed line show the power spectra of the noise for the different maps.}
\label{fig:PS}
\end{figure}

\begin{figure}[htb]
\centerline{\includegraphics[scale=0.35]{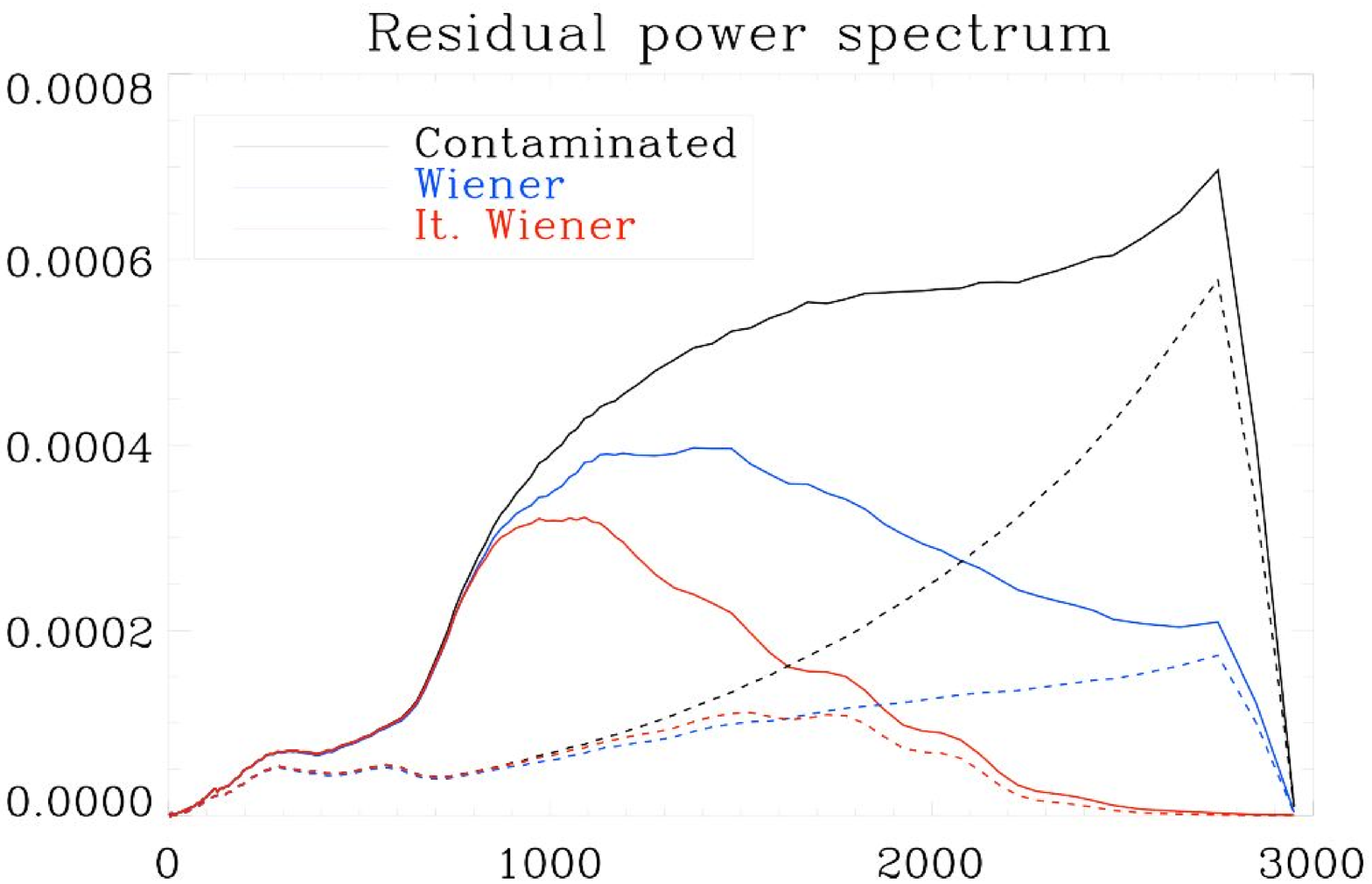}}
\caption{{\bf Residual power spectra.} In abscissa~: spherical harmonics coefficient. In ordinate~: CMB power in $\mbox{mK}^2$. Dashed line show the power spectra of the noise for the different maps.}
\label{fig:ResPS}
\end{figure}

\begin{figure}[htb]
\center{\includegraphics[scale=0.15]{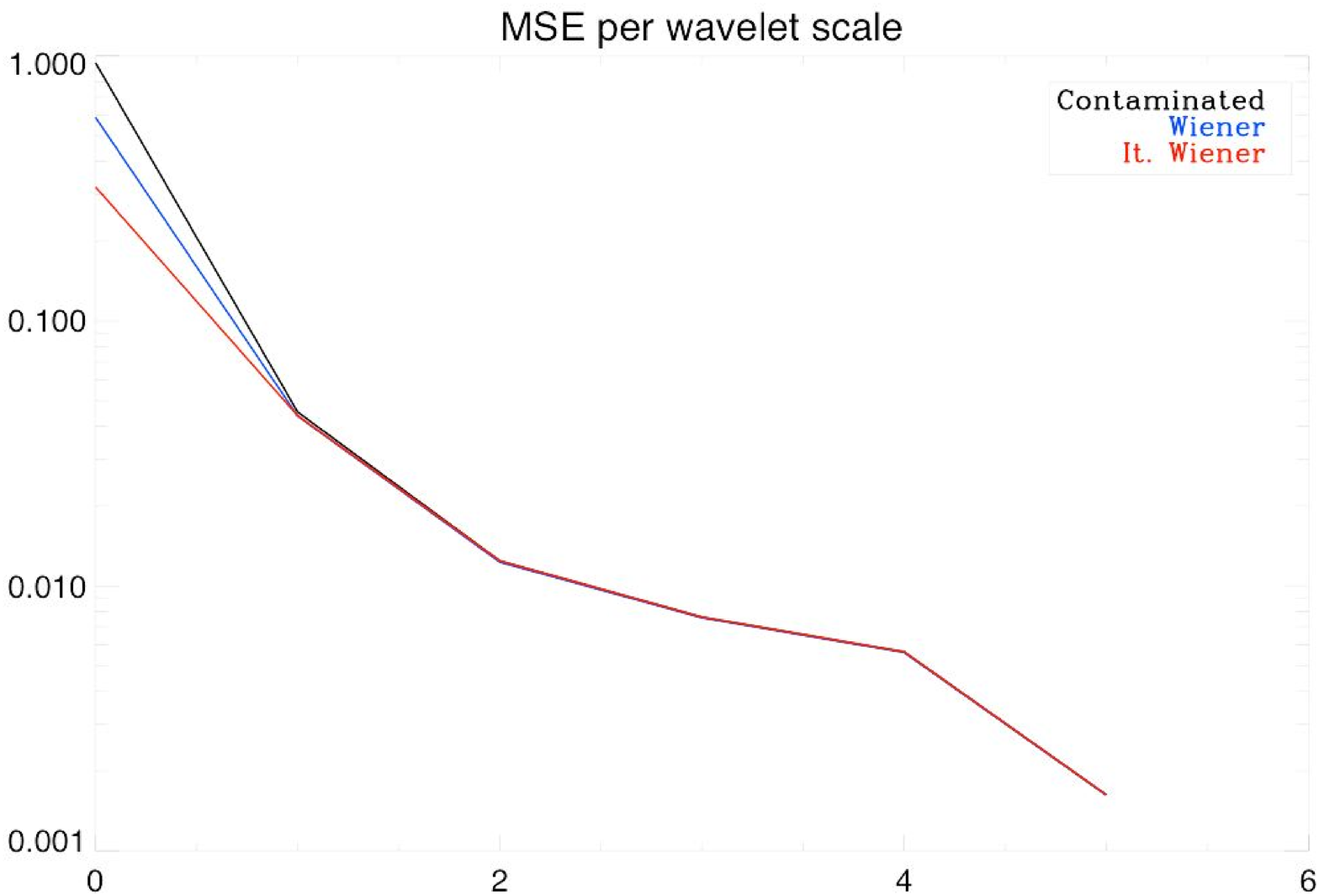}}
\caption{{\bf MSE per wavelet scale.} In abscissa~: wavelet scale - 0 corresponds to $j=1$. In ordinate~: normalized mean squared error.}
\label{fig:MSEWT}
\end{figure}

\begin{figure}[htb]
{\includegraphics[scale=0.25]{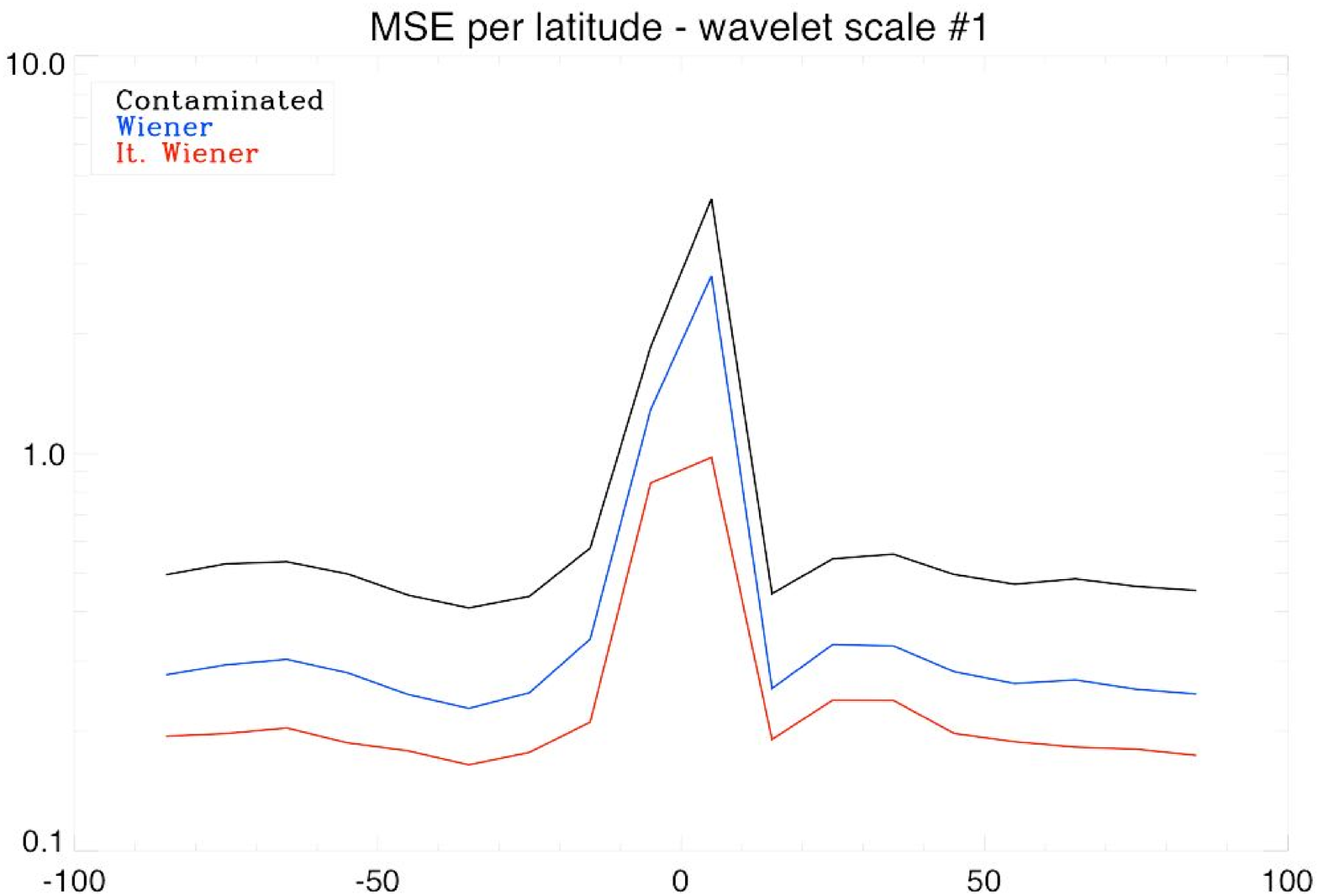}}
\hfill
{\includegraphics[scale=0.25]{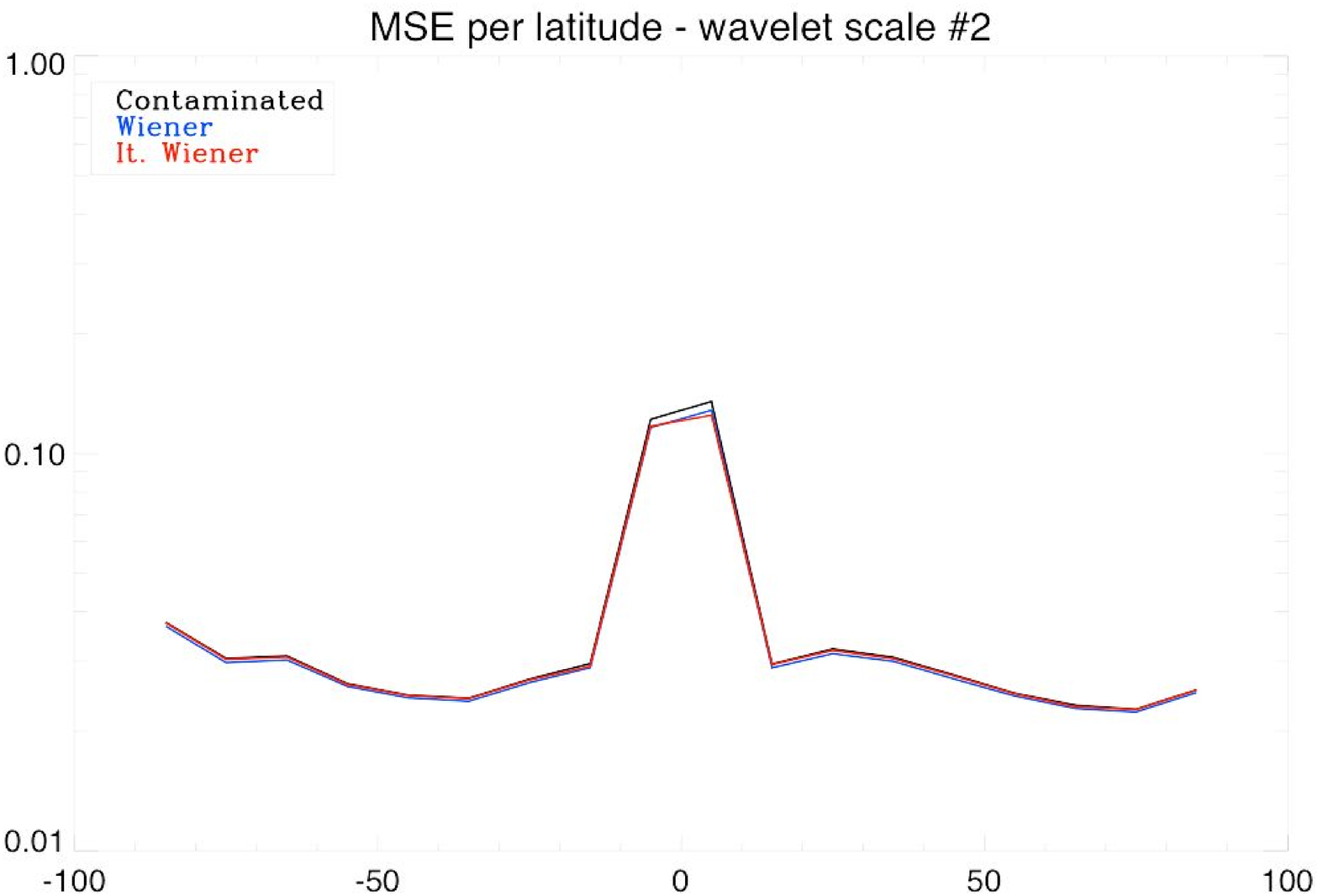}}
\caption{{\bf MSE per latitude at wavelet scale $j=1$ and $j=2$.} In abscissa~: latitude. The galactic plane corresponds to $0^\circ$. In ordinate~: normalized mean squared error.}
\label{fig:MSELat}
\end{figure}

\begin{figure}[htb]
\centerline{\includegraphics[scale=0.5]{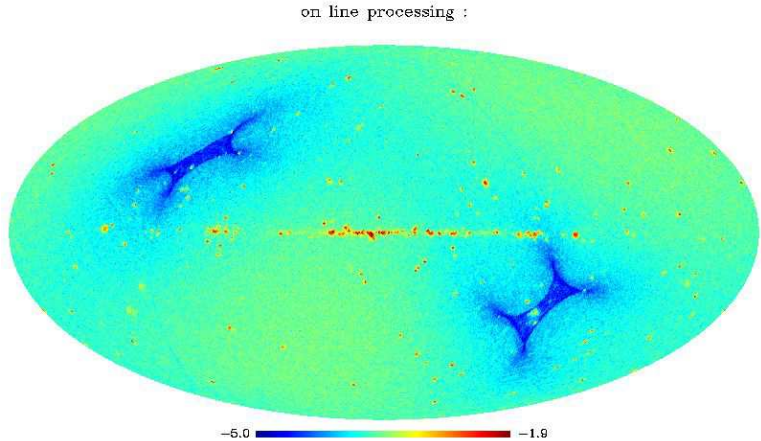}}
\caption{{\bf Estimated noise covariance map at wavelet scale $j=1$~:} this covariance map has been estimated from a single realization of the simulated Planck noise that contaminates the CMB estimated by GMCA. Units in $\mbox{mK}^2$.}
\label{fig:RMSMap_NoFrg}
\end{figure}

In this section, we particularly focus on the noise reduction aspect of the proposed method. The reference denoising method in the field of astrophysics and more specifically CMB data analysis is the \textit{global} Wiener filter applied in spherical harmonics. In opposition to this classical filtering technique, the proposed iterative filtering approach accounts for the non-stationarity of the noise; thereby, it should particularly provide better solution that global Wiener.\\
As detailed in the previous section, the modeling of the noise contribution is performed in the wavelet domain~: the noise is assumed to be non-stationary but decorrelated at each wavelet scale $j$. In each wavelet scale, the noise covariance matrix is  estimated locally from a single noise realization. The noise covariance per wavelet scale is evaluated locally on patches of size $4^{(j+1)}$ where, by convention, $j=1$ is defined as the finest wavelet scale. As an illustration, Figure~\ref{fig:RMSMap_NoFrg} displays the estimated noise RMS map at scale $j=1$.\\
The global Wiener solution is computed by filtering the original CMB map $y$ with the linear filter defined in spherical harmonics by Equation~\eqref{eq:map}. The CMB power spectrum $C_\ell$ is a WMAP7 best-fit estimate; the noise power spectrum $C_\ell^n$ is estimated via the pseudo power spectrum of the noise realization.\\
Iterative Wiener has been applied to the original noisy CMB map $y$. The single parameter to be tuned is the number of iterations $I_{\max}$; in the following experiments, it has been fixed to $250$. Figure~\ref{fig:Maps} shows the noisy map, the LIW-Filtering solution and the noise realization used to estimate the noise local variance in the finest wavelet scale. Figure~\ref{fig:RMaps} displays the residual maps of the {\it global} Wiener and LIW-Filtering solution. As expected the non-stationarity footprint of the noise appears clearly on the residual LIW-Filtering map where more noise seems to have been removed.\\
Figure~\ref{fig:PS} features the power spectra of the original and filtered maps. We would like to point out that these power spectra have been computed from the raw maps; no masking has been applied. This particularly means that these estimates are likely to be biased by strong foreground from the galactic plane. The first observation is that the global Wiener filter does not fully remove the noise at high $\ell$ contrary to the iterative filtering technique. This can be explained by the presence of significant non-stationarities at high $\ell$ which can produce a substantial bias at high frequency. Interestingly, the iterative technique succeed in largely reducing noise at high $\ell$ thus supporting the significant contribution of the noise non-stationarities in producing a bias on the power spectrum. More technically, it is worth noting that the power spectrum of a stochastic process is properly defined as long as this process is wide sense stationary \cite{Chonavel}. Obviously, the non-stationarities of the noise make it depart from this assumption; thereby the power spectrum has no rigorous statistical meaning in this context. Said differently, the power spectrum of the noise $C_\ell^n$ does not contain all the information that a non-stationary noise does in the pixel domain.\\
It is well known that the Wiener filter provides a CMB map that has a biased power spectrum. Thereby the Wiener filter - in red in Figure~\ref{fig:PS} - yield a power spectrum that is negatively biased at high $\ell$  when the noise highly predominates upon the CMB - typically for $\ell > 2500$. \\
Figure~\ref{fig:ResPS} shows the power spectrum of the residual~: $x - \hat{x}$. This quantity is particularly well suited to visualize the power discrepancy in frequency between the original CMB map $x$ and its estimate $\hat{x}$. As expected, the \textit{global} Wiener filtering reduces the contribution of noise at high $\ell$ (see the blue line in Figure~\ref{fig:ResPS}). Still, the filtered map largely departs from the original map. Interestingly, the proposed iterative filtering technique (in red in Figure~\ref{fig:ResPS}) largely outperforms the classical Wiener filter from $\ell = 900$ to $\ell = 3000$. More precisely, the residual power is reduced by up to an order of magnitude for $\ell > 2500$. \\
A more complete measure of discrepancy between the original map $x$ and one of its estimate $\hat{x}$ consists in evaluating the mean squared error (MSE) in the wavelet domain. To that end, Figure~\ref{fig:MSEWT} displays the MSE of different estimators of the CMB map in $5$ wavelet scales in logarithmic scale. The normalized MSE between some CMB map estimate $\hat{x}$ and the ``true" CMB map is computed as follows~: $\frac{\sum_k (\hat{x}[k] - x[k])^2}{\sum_k x[k]^2}$. Again, this figure shows that the iterative technique which directly account for the non-stationarity of the noise, performs better than the \textit{global} Wiener filter. The impact of the noise tends to vanish for larger scales (\textit{typically} for $\ell < 1500$).  \\
To better illustrate the differences between the \textit{global} Wiener filter and its iterative counterparts, Figure~\ref{fig:MSELat} displays the MSE of the different estimators of the CMB map which has been computed in the wavelet domain per bands of latitude of width $10^\circ$. By convention, the galactic plane is centered around the latitude $0^\circ$. The figure on the right displays the MSE which has been evaluated from the finest wavelet scale $j = 1$; this corresponds to an analysis window that ranges from $\ell = 1500$ to $\ell = 3000$. Again, taken globally, one can point out the clear improvement between the {\it global} Wiener filter and its iterative version. As expected the Wiener filter is optimal in the sense of the MSE. More interestingly, the MSE of the two filtering techniques seems to be rather constant for high latitudes and sharply increases in the range $[-15^\circ , 15^\circ]$. As already emphasized, the impact of any of these linear filtering techniques is the more significant when the signal to noise ratio (SNR) decrease. This is obviously the case at smaller scales. At larger scales, the SNR increases and the impact of the filtering is much less significant. This is exactly what can be observed on the plot on the right of Figure~\ref{fig:MSELat}~: the MSE of any of the filtered CMB maps at scale $j=2$, which grossly correspond to an analysis window ranging from $\ell = 750$ to $\ell = 1500$, are approximately the same.\\ \\


As a brief conclusion for these experimental results, it appears clearly that the non-stationarity of the noise has an impact on the CMB estimate. The most classical noise reducing technique in the field, \textit{i.e.} the \textit{global} Wiener filter, only relies on the power spectra of the CMB $C_\ell$ and the noise $C^n_\ell$. Obviously, the behavior of the noise power spectrum contains almost no information about its spatial behavior. These experiments have shown that the noise non-stationary has a clear impact on the CMB. Further accounting for the spatial variations of the noise variance helps improving the noise reduction process by a dramatic amount.

%
%

\section{From noise reduction to contamination reduction}
\label{sec:varestim}
As stated in the introduction in Section~\ref{sec:intro}, one of the purposes of this paper is to tackle the problem of post-separation contamination reduction. In this setting, contamination generally means instrumental noise and residuals of galactic foregrounds. In the case of Planck, CMB maps are generally obtained by applying sources separation methods. Most of these methods provide a solution that can be formulated as a linear combination of the original Planck multichannel observations. Therefore, it is relevant to assume that these contaminants essentially contribute additively to the CMB map~:
\begin{equation}
\label{eq:conteq}
y = x + f + n
\end{equation}
where $f$ stands for the foreground contamination and $n$ for the instrumental noise. Furthermore, it is also natural to assume that these three components $x$, $f$ and $n$ are mutually uncorrelated. The filtering technique introduced in the previous sections mainly relies on the fact that the contaminants are characterized by their covariance matrix $\bf \Sigma$. The questions that we aim at tackling in this paragraph is~: \textit{How far do we know $\bf \Sigma$ and how can it be estimated from the data ?}\\

\subsection{Foreground residuals modeling}

Unlike noise, the foreground residual are generally non-Gaussian, with the exception of the CIB (see \cite{CIB95}). We further restrict the modeling of residuals to their second order statistics thus opting for a Gaussian approximation of their contribution. The contribution of foreground residuals are obviously not known in advance; this means that their covariance matrix has to be estimated from the data directly. From the uncorrelation of the different components, the additive contamination model in Equation~\eqref{eq:conteq} then reads at the level of second statistics~:
$$
{\bf \Sigma} = {\bf \Sigma}_x + {\bf \Sigma}_f + {\bf \Sigma}_n
$$
In this equation, the correlation of foreground pixels implies that ${\bf \Sigma}_f$ is non-diagonal. Similarly to the noise, second order dependencies of foreground pixels are way too complex to be modeled in the pixel domain.\\
So as to capture the correlation of foreground pixels, a natural and simple strategy boils down to adopting the wavelet-based statistical modeling used to model correlated noise in the previous Section~\ref{sec:wtmodel}. To this end, $w^f_{j}[k]$ will denote the wavelet coefficient of $f$ in scale $j$ at pixel $k$ with the convention~: $j=1$ corresponds to the finest scale and $j=J-1$ to the coarse resolution. From this definition, we will assume that these wavelet coefficients verify~:
\begin{equation}
\mathbb{E}\left \{ w^f_{j}[k] w^f_j[i] \right \} = {\sigma_{j}^f}[k]^2 \delta_{k,i}
\end{equation}
where ${\sigma_{j}^f}[k]^2$ stands for the local variance of the foreground at pixel $k$. The covariance matrix of the foreground residual in wavelet scale $j$ is diagonal and equal to~:
\begin{equation}
{\bf \Sigma}_{f,j} = \mbox{diag}\left({\sigma_{j}^f}[k]^2\right)
\end{equation}
In practical situations, ${\bf \Sigma}_{f,j}$ has to be estimated from the data $y$ themselves. Similarly to the modeling of noise, the local noise variance of the foreground will be approximated by the empirical estimate of the variance in a surrounding neighborhood of size $\sqrt{P}\times \sqrt{P}$ centered about the pixel $k$~:
\begin{equation}
\label{eq:varest}
\hat{\sigma}_{j}^f[k]^2 = \frac{1}{P} \sum_{i \in \mathcal{N}[k]} {w^f_{j}}[i]^2
\end{equation}

\subsection{Contamination filtering}

In the remaining of this paper, we will generally use the term contamination for the contribution of both the foreground residuals and the instrumental noise. In the following, as suggested previously, we also opt for exactly the same model to capture the pixel dependencies of the instrumental noise. Therefore, we will now define locally at pixel $k$ and scale $j$ the variance of contamination (noise and foreground) as $\hat{\sigma}_{j}^c[k]^2 = \hat{\sigma}_{j}^f[k]^2 + \hat{\sigma}_{j}^n[k]^2$. The joint contribution of noise and foreground residuals will be evaluated jointly from the data.\\

From the estimation of the estimation of the local variance $\hat{\sigma}_{j}^y[k]^2$; one has to recall that the contaminants' variance is not directly accessible but is equal to~:
\begin{equation}
\hat{\sigma}_{j}^c[k]^2 = \left [ \hat{\sigma}_{j}^y[k]^2 - \hat{\sigma}_{j}^2\right ]_+
\end{equation}
where $ \hat{\sigma}_{j}^2$ stands for the variance of the CMB and $[ x ]_+ = \max (0,x)$. Let use note that, from the stationarity of the CMB map, $\hat{\sigma}_{j}^2$ does not explicitly depend on $k$; this holds as long as the patch size at scale $j$ is large enough compared to the CMB fluctuations (\textit{i.e.} correlation length) in the $j$-th wavelet scale. The value of the CMB variance at scale $j$-th is directly related to the CMB power within this scale; its value can be derived from the CMB power spectrum $C_\ell$ and the wavelet analysis filters. To that end, let us denote by $\psi_{j,\ell}$ the spherical harmonics filter from which the wavelet scale $j$ is defined. As emphasized in Section~\ref{sec:intro} the spherical harmonics coefficients of the CMB map  should be mutually uncorrelated. This particularly entails that the power of the CMB in the $j$-th wavelet scale is exactly~:
$$
\sigma_j^2 = \frac{1}{4 \pi}\sum_\ell (2 \ell + 1) \psi_{j,\ell}^2 C_{\ell}
$$
It follows that the variance of the contamination at pixel $k$ is approximately equal to~:
\begin{equation}
\label{eq:locvar}
\hat{\sigma}_{j}^c[k]^2 = \left [ \hat{\sigma}_{j}^y[k]^2 - \frac{1}{4 \pi}\sum_\ell (2 \ell + 1) \psi_{j,\ell}^2 C_{\ell} \right ]_+
\end{equation}
Thereby, from the \textit{a priori} knowledge of the CMB power spectrum $C_\ell$, an approximate contribution of the residual can be evaluated at each pixel and each wavelet scale. The reader has to keep in mind that the measurement of this contribution is only approximate and has to be considered under the light of the following assumptions~: 1) the contribution of the residual is modeled as a non-stationary Gaussian field where the covariance matrix is diagonal in each wavelet scales and 2) its variance being evaluated on small patches of size $\sqrt{P} \times \sqrt{P}$, it is assumed to be approximately stationary in a given analysis patches. However, even with these approximations, our model offers a much more realistic modeling of the complexity of the data than the traditional Wiener method. Improvements of this modeling will be discussed in Section~\ref{sec:discussion}.\\
Let ${\bf \Sigma}_{f,j} = \mbox{diag}(\hat{\sigma}_{j}^c[k]^2)$. Then, in the algorithm introduced in Section~\ref{sec:filtering}, the noise matrix $\bf \Sigma$ is substituted with the following covariance matrix~:
$$
{\bf \Sigma} = W^T \left [
\begin{array}{ccc}
{\bf \Sigma}_{f,1} & { \bf 0} & {\bf 0} \\
{\bf 0} & \ddots & {\bf 0} \\
{\bf 0} & {\bf 0} & {\bf \Sigma}_{f,J} \\
\end{array}
\right ]W
$$
where ${\bf W}$ stands for the isotropic wavelet transform and ${\bf W}^T$ its adjoint.
Along with the iterative technique introduced in Section~\ref{sec:filtering}, the resulting wavelet-based filtering will be coined LIW-Filtering for Linear Iterative Wiener Filtering in the following.

\subsection{Contamination reduction on Planck simulated data}
\label{sec:contreduc}

\begin{figure}[htb]
\centerline{\includegraphics[scale=0.5]{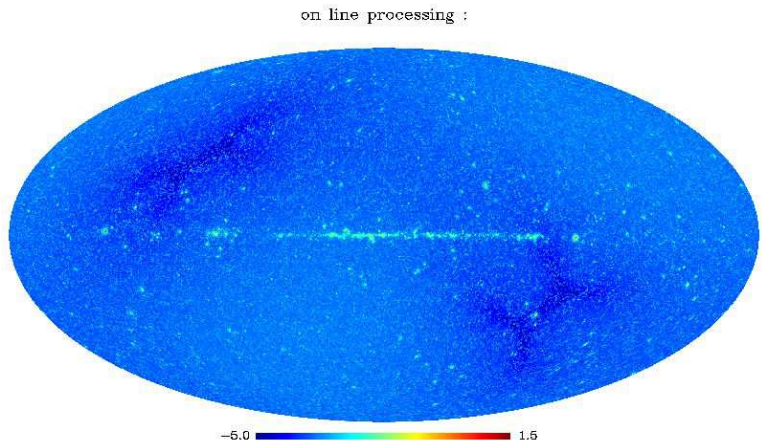}}
\caption{{\bf Estimated contamination covariance map at wavelet scale $j=1$~:} this covariance map has been estimated from the CMB estimated by GMCA. Units in $\mbox{mK}^2$.}
\label{fig:RMSMap_Frg}
\end{figure}

\begin{figure}[htb]
\centerline{\includegraphics[scale=0.5]{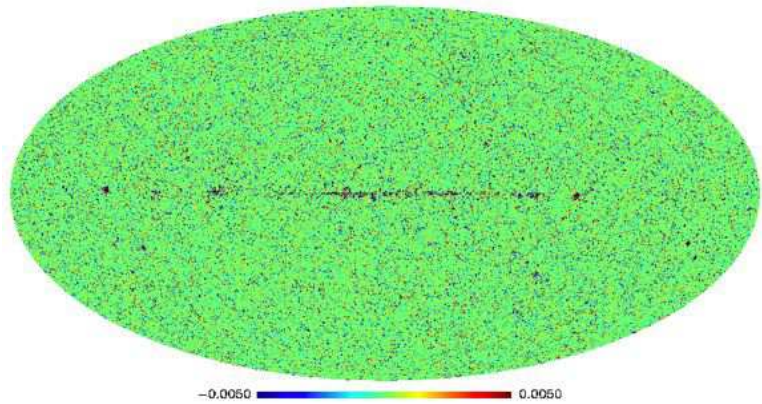}}
\caption{{\bf CMB and residual maps} Difference map between the noise-only LIW-Filtering solution and the contamination-aware LIW-Filtering solution in the finest wavelet scale. Units in $\mbox{mK}$.}
\label{fig:RMaps2}
\end{figure}

\begin{figure}[htb]
\centerline{\includegraphics[scale=0.35]{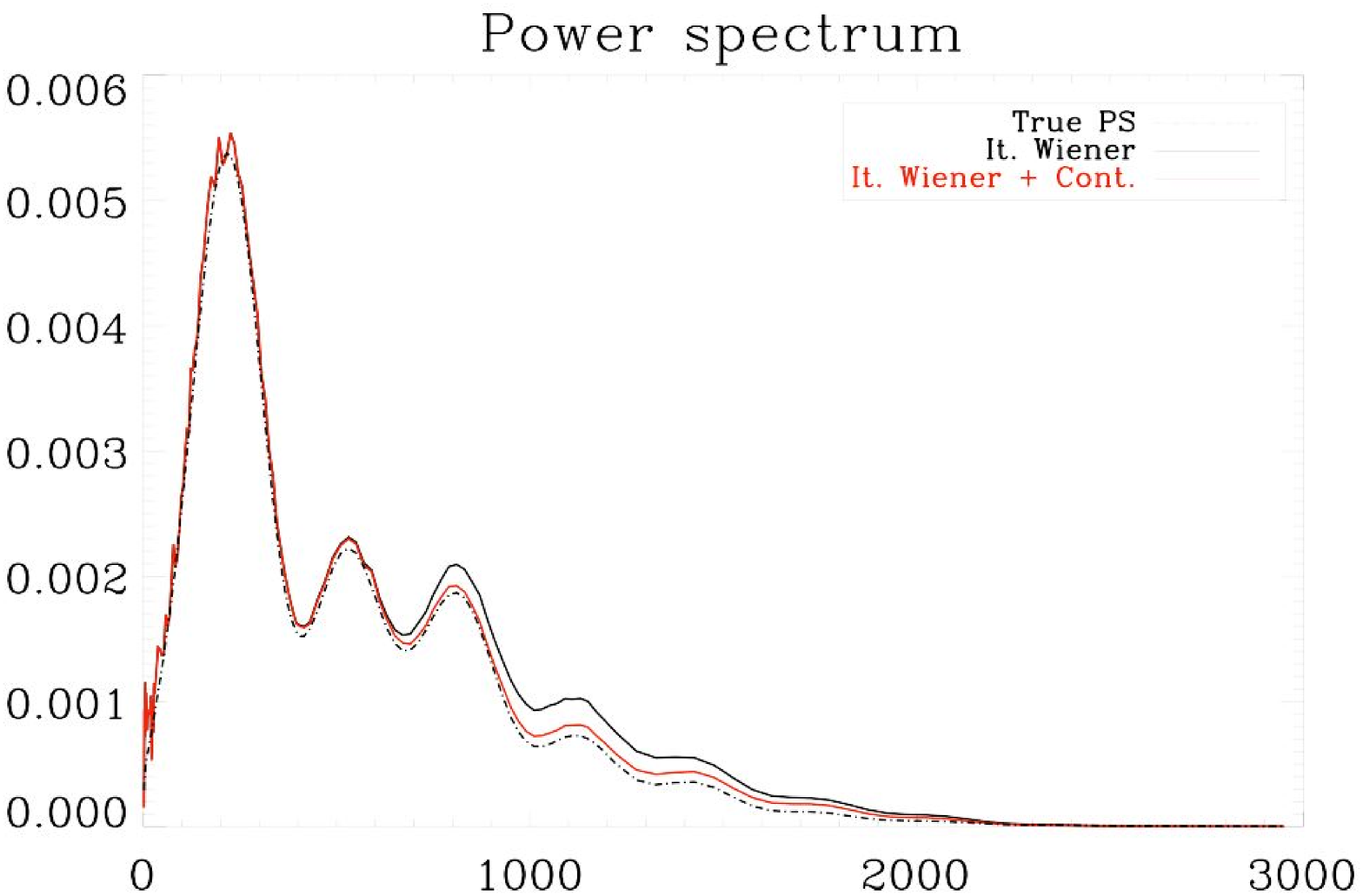}}
\caption{{\bf Power spectra.}. In abscissa~: spherical harmonics coefficient. In ordinate~: CMB power in $\mbox{mK}^2$. Dashed line show the power spectra of the noise for the different maps.}
\label{fig:PS_Frg}
\end{figure}

\begin{figure}[htb]
\centerline{\includegraphics[scale=0.35]{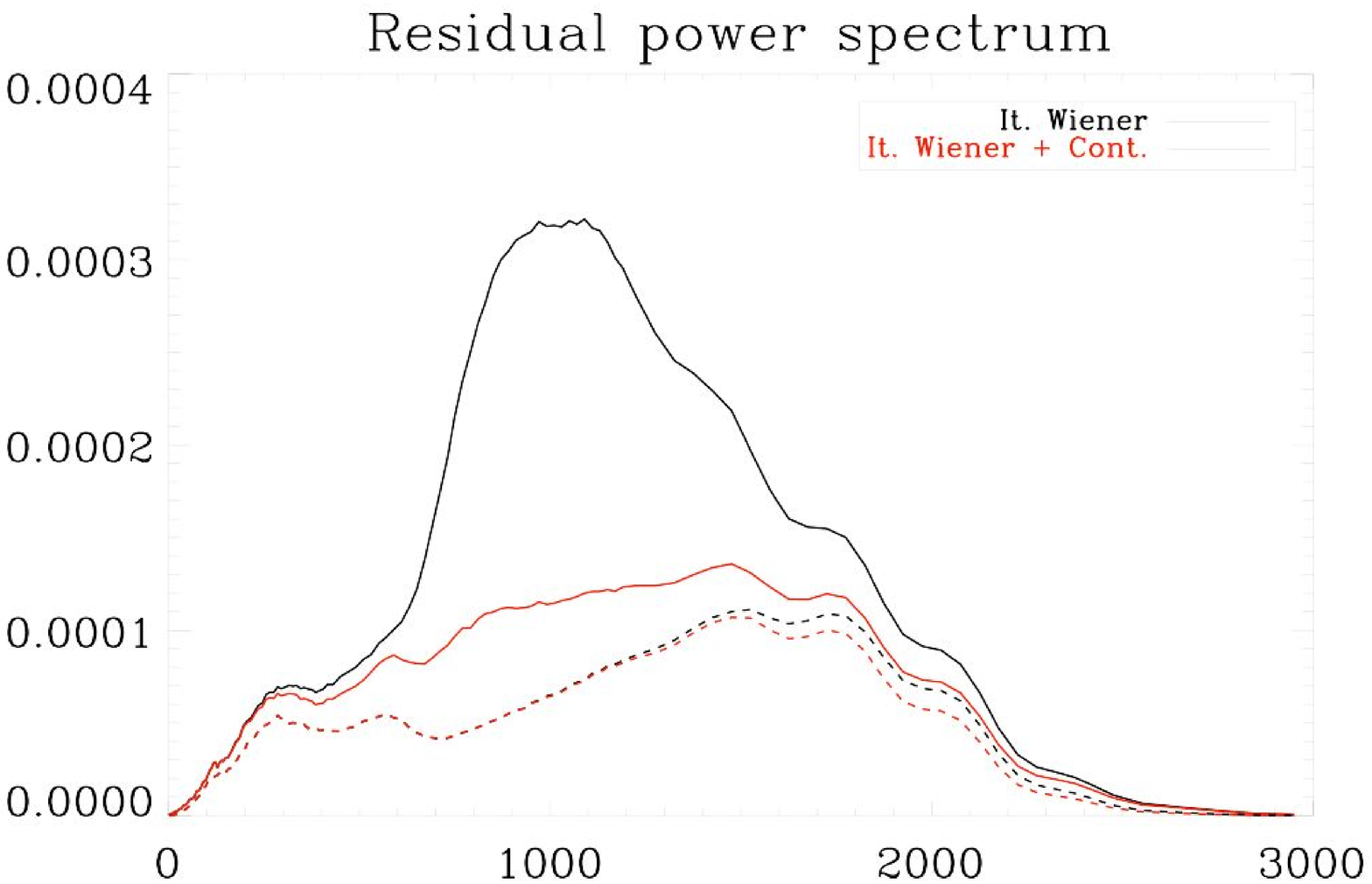}}
\caption{{\bf Residual power spectra.} In abscissa~: spherical harmonics coefficient. In ordinate~: CMB power in $\mbox{mK}^2$. Dashed line show the power spectra of the noise for the different maps.}
\label{fig:ResPS_Frg}
\end{figure}

\begin{figure}[htb]
\center{\includegraphics[scale=0.35]{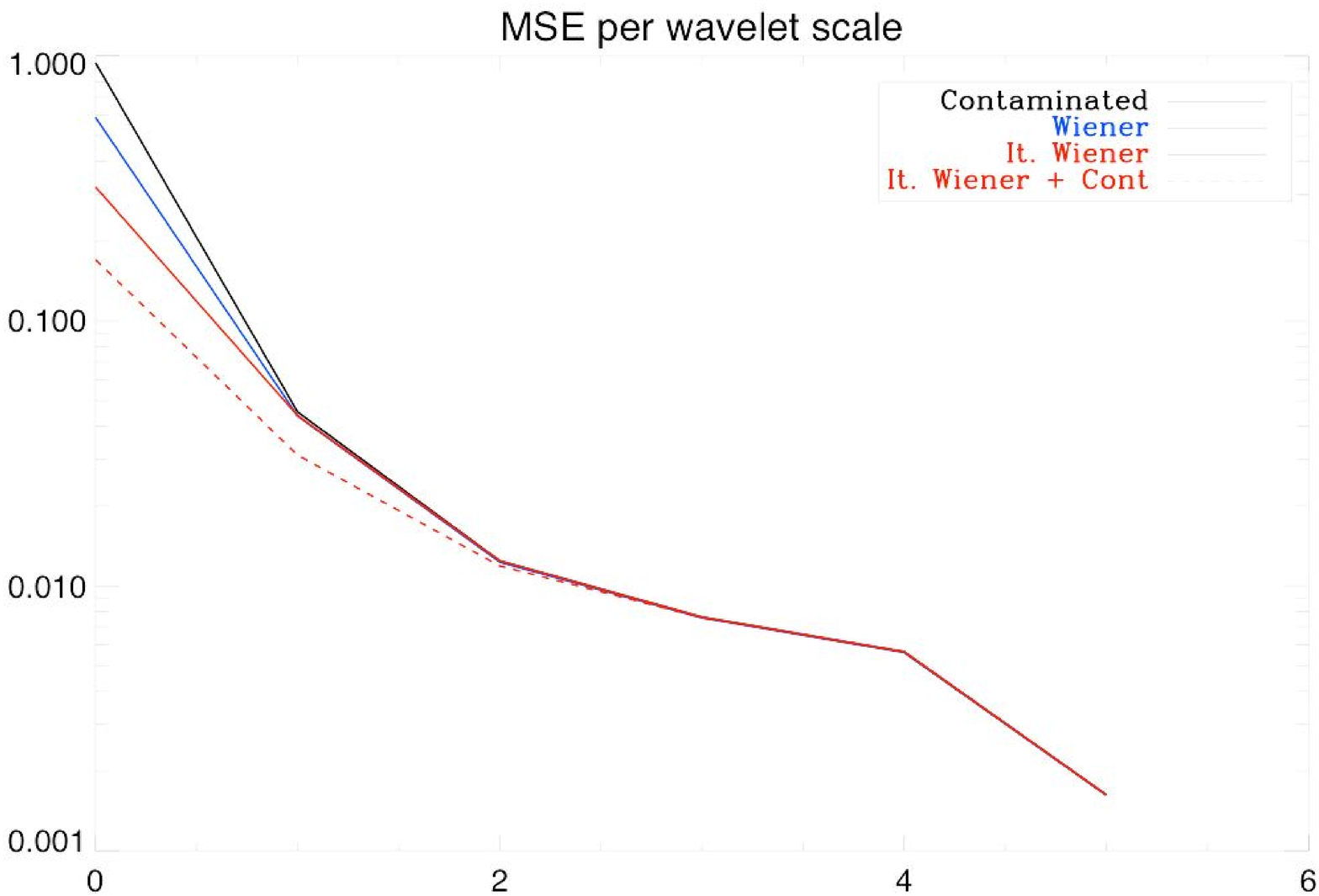}}
\caption{{\bf Normalized MSE per wavelet scale.} In abscissa~: wavelet scale - 0 corresponds to $j=1$. In ordinate~: normalized mean squared error.}
\label{fig:MSEWT_Frg}
\end{figure}

\begin{figure}[htb]
{\includegraphics[scale=0.25]{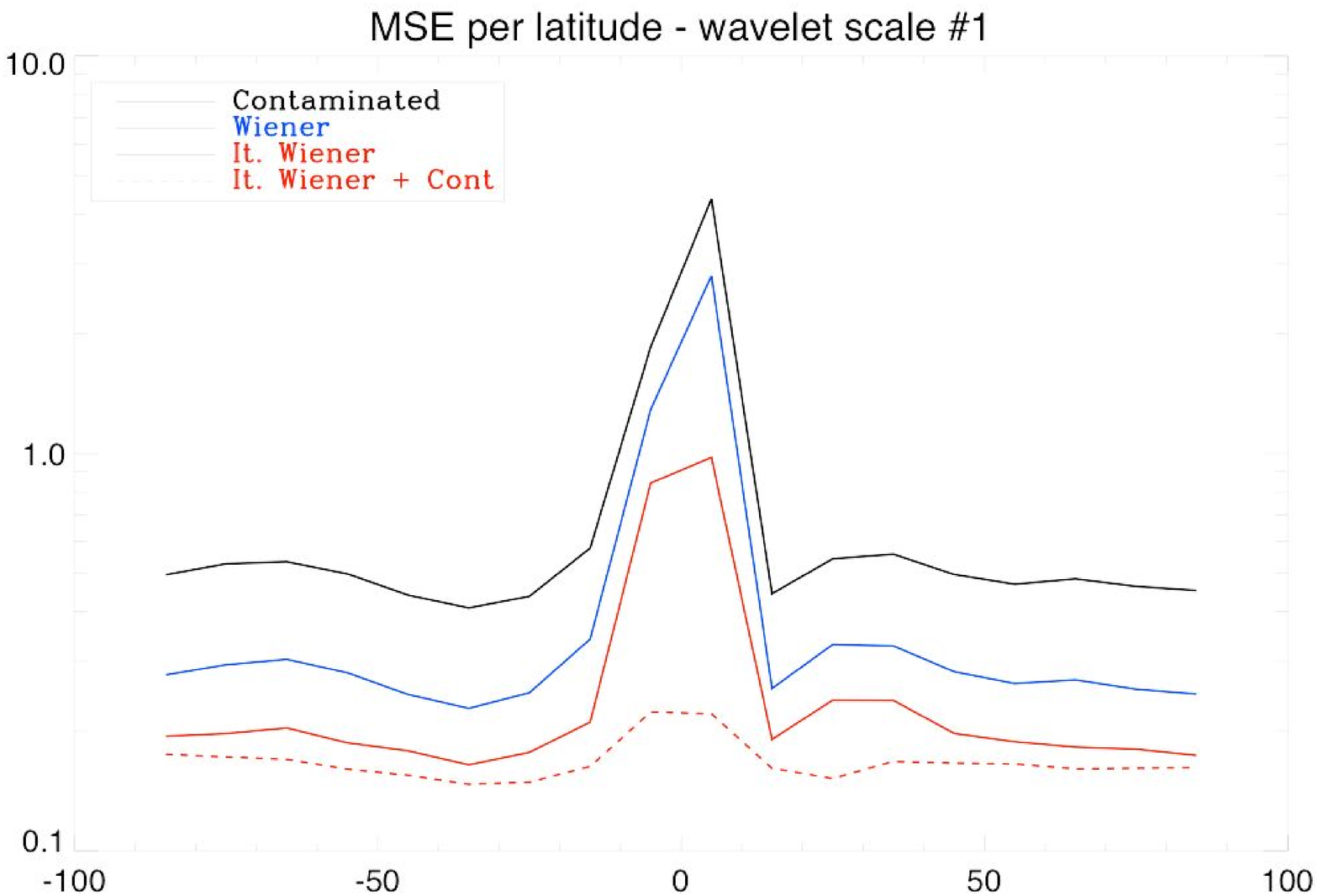}}
\hfill
{\includegraphics[scale=0.25]{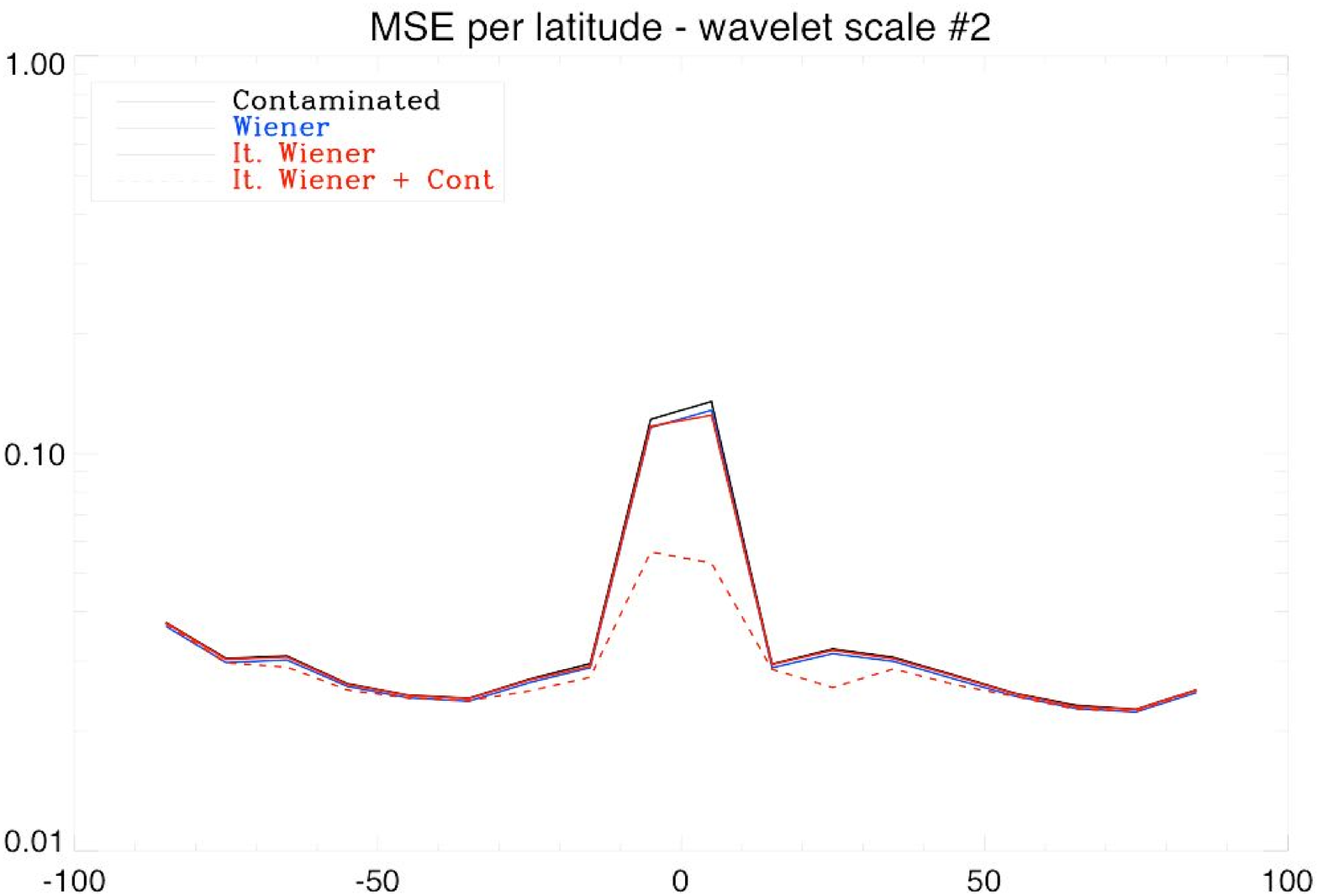}}
\caption{{\bf Normalized MSE per latitude at wavelet scale $j=1$ and $j=2$.} In abscissa~: latitude. The galactic plane corresponds to $0^\circ$. In ordinate~: normalized mean squared error.}
\label{fig:MSELat_Frg}
\end{figure}

\begin{figure}
\center{\includegraphics[scale=0.35]{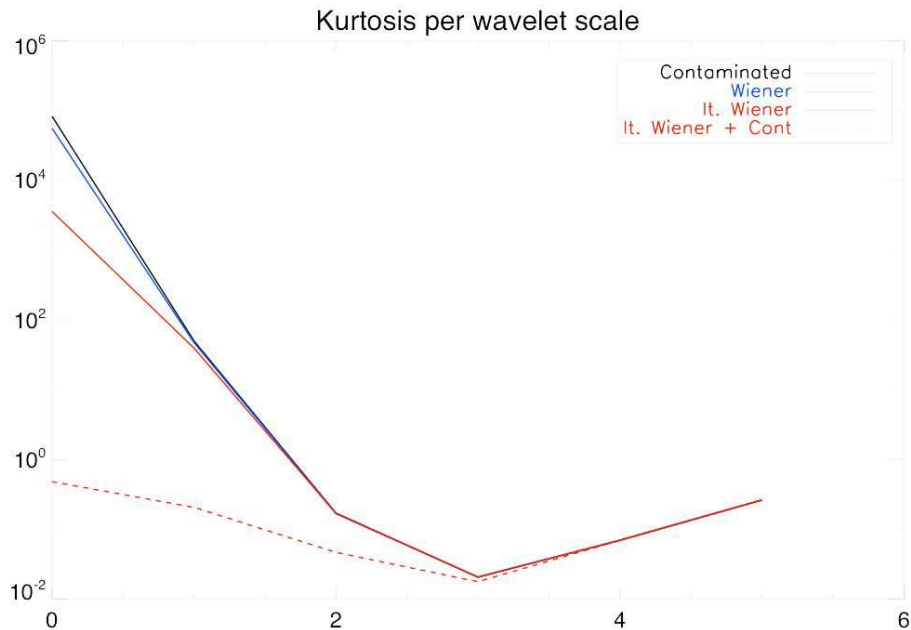}}
\caption{{\bf Kurtosis $\kappa_4$ per wavelet scale.} In abscissa~: wavelet scale - 0 corresponds to $j=1$. In ordinate~: value of the kurtosis $\kappa_4$.}
\label{fig:KurtWT}
\end{figure}

\begin{figure}[htb]
{\includegraphics[scale=0.25]{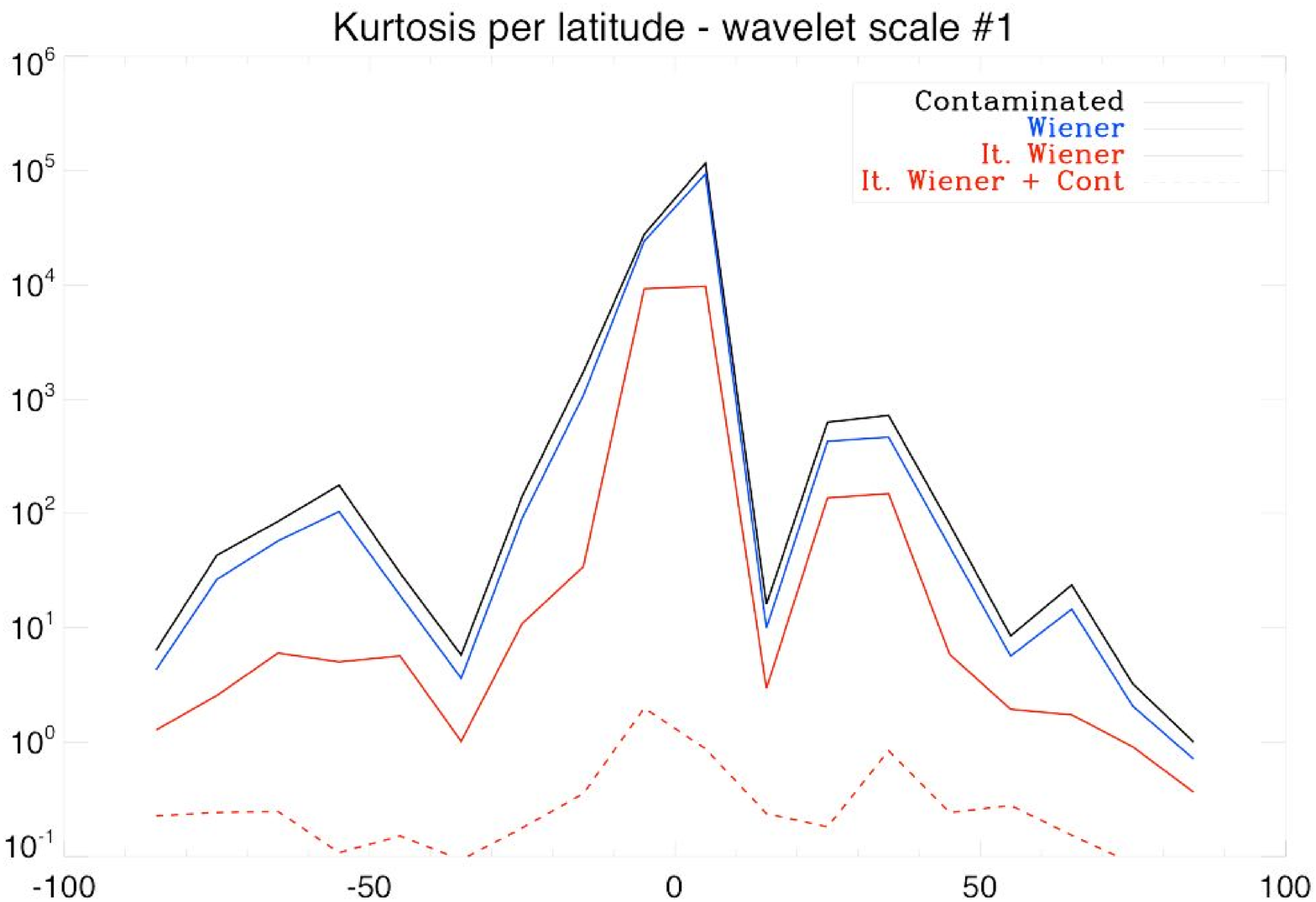}}
\hfill
{\includegraphics[scale=0.25]{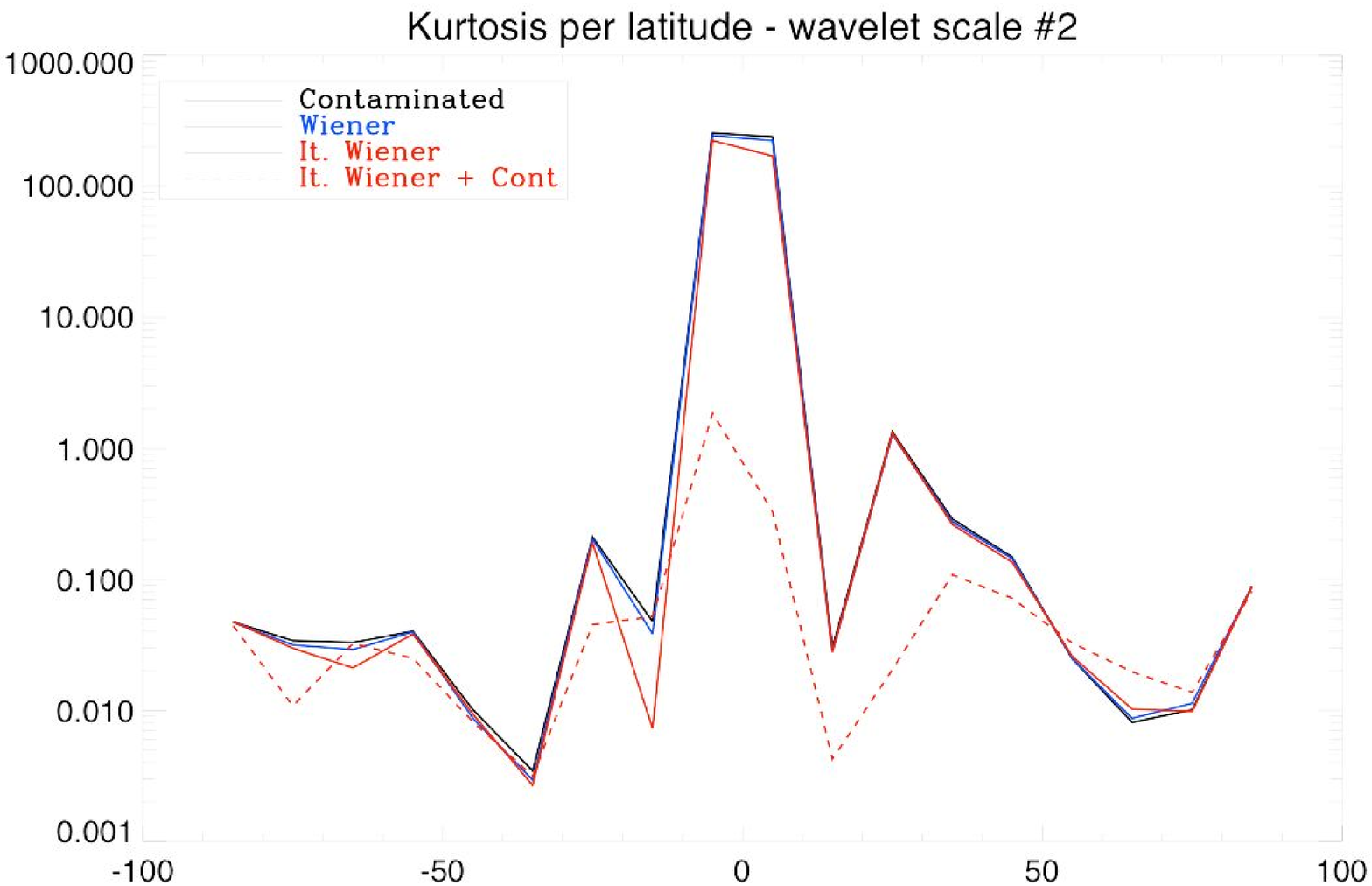}}
\centerline{\includegraphics[scale=0.25]{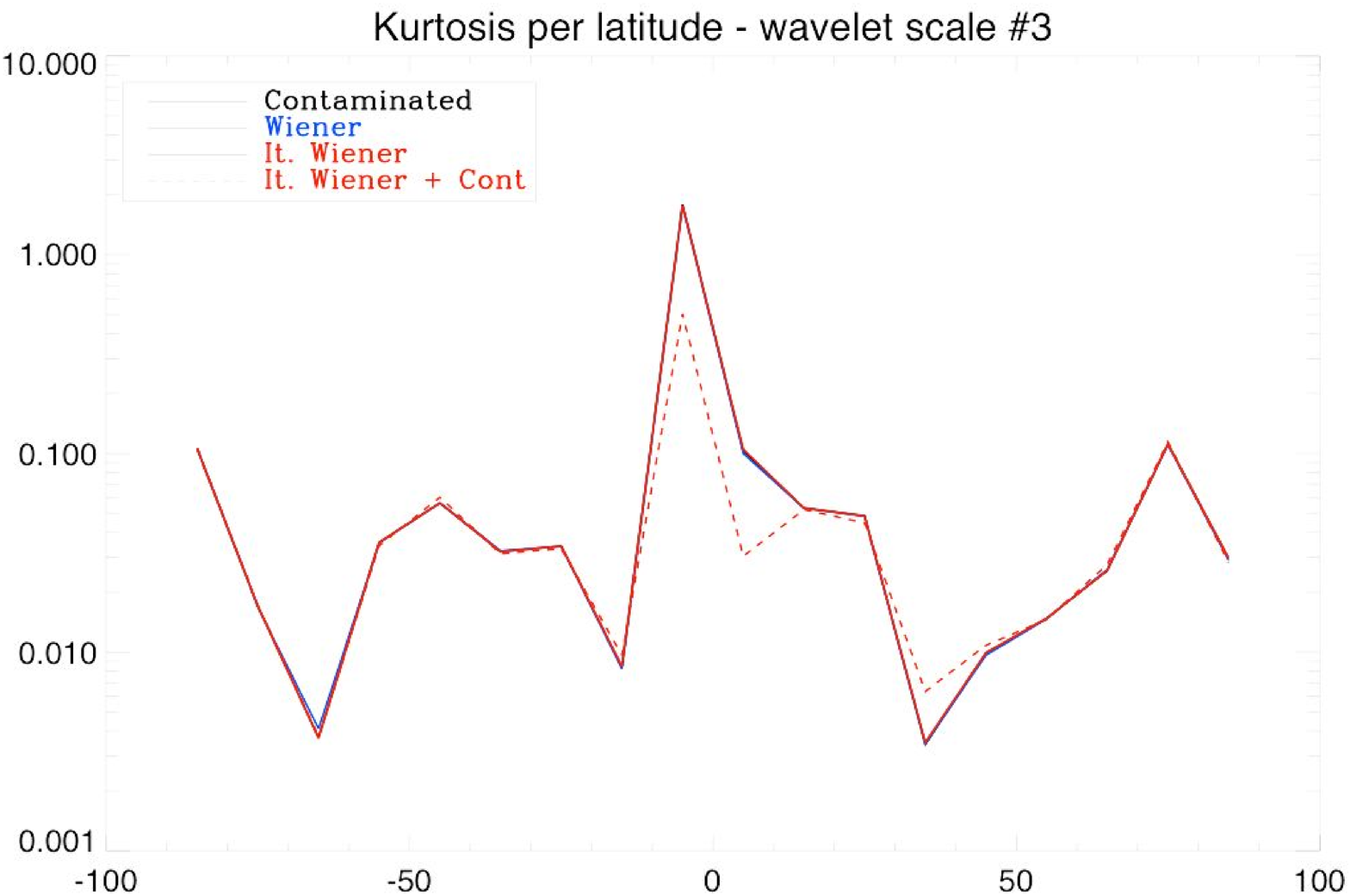}}
\caption{{\bf Kurtosis $\kappa_4$  per latitude at wavelet scale $j=1$ and $j=2$.} In abscissa~: latitude. The galactic plane corresponds to $0^\circ$. In ordinate~:  value of the kurtosis $\kappa_4$.}
\label{fig:KurtLat}
\end{figure}

As emphasized in Section~\ref{sec:varestim}, the wavelet-based noise/contamination variance modeling used so far makes it possible to account for the contribution of the foreground contamination in the filtering process. Estimating the contribution of the foreground contamination, as detailed in Section~\ref{sec:varestim}, makes the reduction of contamination possible. In the following experiments, the contamination, which includes the contribution of the noise and an approximate contribution of the foreground residuals, has been modeled in the wavelet domain \cite{starck:sta05_2} using $6$ scales. Within each wavelet scale, with the exception of the coarsest scale to which no filtering is applied, the contribution of the contamination is assumed to be Gaussian and locally stationary on patches of size $\left \{ 4^{j+1}\right \}_{j=1,\cdots,5}$. Figure~\ref{fig:RMSMap_Frg} features the estimated variance map in logarithmic scale. As a remark, if the morphology of this contaminant variance map resembles at first glance the noise-only variance map, the dynamic range of the former is clearly larger by $3$ orders of magnitudes. These differences are mostly significant on the galactic plane and on local features that are likely point sources or residuals of SZ clusters.\\
The proposed Wiener based iterative method has been applied on the aforementioned Planck simulated data. Figure~\ref{fig:RMaps2} shows the difference map between the noise-only LIW-Filtering solution and the proposed contamination-aware LIW-Filtering solution in the finest wavelet scale. It appears clearly that the difference between the map are mainly concentrated on the galactic plane and on likely point sources at high latitude.\\
Figure~\ref{fig:PS_Frg} displays the power spectra of the original and filtered CMB maps along with the theoretical power spectrum of the simulated map in black dash-dotted line. At first glance, accounting for the contribution of the foreground residuals, even approximately, yields improvements from $\ell = 750$. Perhaps more interesting, Figure~\ref{fig:ResPS_Frg} shows the power spectrum of the residual maps $x - \hat{x}$~: the contaminant-aware filtering technique clearly outperforms the previous methods from $\ell = 250$ to $\ell = 3000$ with a dramatic improvement in the range $[750,1500]$.\\
Figure~\ref{fig:MSEWT_Frg} displays the wavelet-based MSE we described in the previous paragraph. The results of the contaminant-aware filtering technique have been further added to the plot in Figure~\ref{fig:MSEWT}. Let us recall that the noise reducing techniques described in the previous section mainly help reducing the noise in the noise-dominant regime. This particularly explained why the MSE remains almost unaltered in wavelet scales $j > 1$. Interestingly, accounting for an approximate contribution of the foregrounds helps reducing their impact in larger scales, typically for $j < 3$.\\
Figure~\ref{fig:MSELat_Frg} presents the MSE computed at different latitudes in the finest wavelet scale $j=1$ - plot on the left - and $j=2$ -plot on the right. Accounting for the contaminant contribution improves the MSE of the filtered map at all latitudes. Interestingly, the MSE is decreased by a large amount - a factor of approximately $5$ -  on the galactic (latitude $0^\circ$). The same holds in a slighter amount in wavelet scale $j=2$.\\
Assuming that the Planck instrumental noise is Gaussian is widely considered as a reasonable assumption. This is obviously not the case for foreground residuals the presence of which is likely to largely distort the search for non-gaussian features in the CMB. Thereby, reducing the amount of contaminant should help preventing non-gaussianity test from the non-Gaussian impact of foreground residuals. In the field of CMB non-gaussianity evaluation, a classical technique boils down to measuring higher order statistics in the wavelet domain \cite{starck:sta03_1}. One of the statistics of choice is the statistics of order $4$, denoted $\kappa_4$, a.k.a. the kurtosis. The kurtosis of the wavelet coefficients of the estimated CMB map has been evaluated in $5$ scales; non-gaussianity test results are shown in Figure~\ref{fig:KurtWT}. The first observation is that the proposed method helps reducing non-gaussian contamination even when only the noise is modeled as a spurious component. This suggests that the non-stationary behavior of the noise can generate undesired non-gaussian signatures. Thereby, accounting for the non-stationary behavior of the noise in the noise reduction process decreases the kurtosis of the filtered map by an order of magnitude in the finest wavelet scale.\\
The red dashed line in Figure~\ref{fig:KurtWT} shows a clear reduction of the kurtosis of the filtered map when the contribution of the foreground residuals is modeled. More precisely, the value of $\kappa_4$ is grossly reduced by $4$ orders of magnitude for $j=1$, $2$ orders of magnitude for $j=2$ and $1$ order of magnitude for $j=3$. While noise-only filtering yields map that departs from the Gaussian assumption for $J < 2$, the contamination-aware filtered map is now compatible with the Gaussian assumption in all wavelet scales with $\kappa_4 < 1$.\\
Figure~\ref{fig:KurtLat} presents more detailed non-gaussianity tests; the three plots that this figure displays show the evolution of $\kappa_4$, in the three finest wavelet scales $j=1,\cdots,3$, which has been evaluated in bands of latitude of width $10^\circ$. The galactic plane is centered about the point $0^\circ$. The non-Gaussian contamination mainly originates from the presence of galactic foreground residuals which are mainly concentrated on the galactic plane. This explains the very large value of $\kappa_4$ in the range of latitude $[-15^\circ,15^\circ]$. The second well-known source of non-Gaussian signatures at spatial high frequency are the radio and infrared point sources. These foregrounds are roughly uniformly scattered all over the sky. Point sources are likely to be the predominant source of non-Gaussian signatures at high latitude thus explaining the relatively large value of $\kappa_4$ in this range of latitudes. As shown in the top-left plot of Figure~\ref{fig:KurtLat}, further filtering the contamination (see the red dashed line) yields a dramatic reduction of $\kappa_4$ up to $5$ orders of magnitude on the galactic plane. The filtered CMB map is likely to be compatible with the Gaussian assumption at every latitude, even on the galactic plane. In the second wavelet scale $j=2$, the same reduction of non-Gaussian signatures is also visible by a lesser extent, mainly on the galactic plane in the range $[-15^\circ,15^\circ]$. In the third wavelet scale, the proposed filtering is likely to reduce the non-Gaussian signatures in the galactic plane; in this range of scales (\textit{i.e.} the third wavelet scale roughly corresponds to an analysis window that spans the range $[375,750]$), the CMB is largely predominant which entails that~: i) the input map is already quite compatible with the Gaussian assumption and ii) the impact of the filtering may be less visible when non-Gaussianity is measured.

\section{Discussion, prospects and conclusion}
\subsection{Discussion and prospects}
\label{sec:discussion}

\paragraph{Filtering the maps, for what use ?}
Important cosmological tests and evaluations are performed on estimated CMB maps; this is particularly the case for CMB lensing reconstruction \cite{CMBLens} and non-Gaussian signature detection to only name two. These tests are particularly sensitive to spurious components that contaminate the estimated CMB map, whether it is noise or foreground residuals. This is therefore crucial to estimate a CMB map that is the less contaminated possible. The second point we would like to emphasize is that the aforementioned cosmological tests or evaluations are generally performed on the full-sky data; prior to any post-processing the estimated map is masked to get rid of the impact of galactic foregrounds and point sources. This has two major consequences~: i) it limits the size of the samples to which these tests/evaluations are performed thus increasing the statistical variance of these tests and ii) any of these tests has to dealing with this mask (\textit{e.g.} via an inpainting procedure in \cite{CMBLens}) thus making the analysis of the CMB map generally more complex.\\
As shown in the previous experiments, the proposed filtering clearly limits the impact of noise by taking into account its non-stationary behavior. Furthermore, results are presented that clearly show that modeling the contribution of foreground residuals make contamination reduction effective even on the galactic plane. This has two major consequences~: i) it makes it possible the use of a much smaller mask prior to any analysis of the estimated CMB map and ii) it helps reducing the non-Gaussian features that originate from the presence of foreground residuals.\\

\paragraph{Beyond the proposed methods} Whether noise or foreground residual is at stake, the central assumption that is at the heart of the modeling of contamination is the decorrelation hypothesis we made in Section~\ref{sec:varestim}. It is assumed that the contaminants have potentially non-stationary but decorrelated samples in each wavelet scales. The basic idea that supports this assumption is the well-known decorrelating property of wavelets bases. If this assumption almost holds for rather theoretical stochastic processes (\textit{e.g.} fractional Brownian motion \cite{fBM92} to only name one), this is only roughly approximate for more general signals such as galactic foreground or complex correlated noise.\\
The basic idea behind this assumption is that the multiscale analysis basis is able to capture the \textit{correlation morphology} of the component to be modeled. In the context of Planck, it is well understood that instrumental noise will be highly correlated along the scanning direction. This means that the correlation of the Planck instrumental noise be better modeled as a decorrelated stochastic process in a multiscale signal representation that best represent elongated structures such as the Curvelet tight frame \cite{starck:sta05_2}. We can extrapolate the same argument to the modeling of foreground residuals.\\
The contamination modeling used so far in this article makes profit of the decorrelation assumption; it particularly help simplifying the filtering process by only requiring the handling of diagonal matrices (\textit{i.e.} root mean squared maps). Departing from the decorrelation assumption will largely increase the complexity of the proposed filtering techniques. However, a straightforward way of extending the proposed methods is to choose multiscale signal representations which are better adapted to the morphology or structure of the signal to be modeled.\\

\paragraph{Potential impact of contamination reduction on CMB non-gaussianities} It is important to wonder whether the contamination reduction filtering introduced in Section~\ref{sec:varestim} might affect or reduce potential CMB-related non-Gaussianities. One important point is that the way the local variance of the contamination is estimated in Section~\ref{sec:varestim} leads, in practice, to an estimation of foreground contaminations that largely exceeds the average CMB power in each wavelet scale. Therefore, the filtering should have a much lesser impact on non-gaussianities whose magnitude is of the order or lower than the magnitude of the CMB. It is worth pointing out that the rule in Equation~\ref{eq:locvar} to estimate the local contamination variance can also be chosen differently to only account for non-gaussianities that are guaranteed to largely exceed the magnitude of the CMB thus avoiding for sure any effect on potential CMB-related non-gaussianities. Future work will focus on evaluating the potential impact of the proposed filtering technique on CMB lensing reconstruction and non-gaussianity detection.

\subsection{Conclusion}
\label{sec:conclusion}
Cosmological microwave background maps estimated from full-sky surveys such as WMAP or more recently Planck generally suffers from various sources of contaminations~: i) instrumental noise is generally non-stationary which may generate non-gaussian signatures and ii) foreground residuals generally remain even after the application of state-of-the-art source separation methods. In this context, the most classical denoising technique, aka. {\it global} Wiener filter; despite its simplicity, it is not able to account for the non-stationarity of the instrumental noise. To that end, we introduce a novel noise reduction technique coined LIW-Filtering for Linear Iterative Wavelet Filtering which combines the linearity of the \textit{global} Wiener filter while accounting for the potential non-stationarity of the noise. In this framework, the noise is modeled as a non-stationary but decorrelated process in the wavelet domain. The denoising problem then takes the very simple form of a quadratically regularized least square where the noise covariance matrix is diagonal in the wavelet domain and the signal covariance matrix is also diagonal in the spherical harmonic domain (\textit{i.e.} CMB power spectrum). The solution is computed using recently introduced proximal algorithms. When noise reduction is at stake, we showed that the proposed iterative technique succeeds in reducing the mean squared error (MSE) of the filtered solution with respect to the \textit{global} Wiener filter classically applied in the field. Moreover, the modeling/estimation framework introduced in this articles makes the reduction of foreground contamination possible. Similarly to the non-stationary instrumental noise, foreground contamination are modeled as a non-stationary but decorrelated process in the wavelet domain. Numerical experiments show that~: i) the MSE of the filtered map is improved; specifically on the galactic plane and 2) very interestingly, non-Gaussian signatures are also dramatically reduced. This particularly provides arguments supporting the application of the proposed filtering technique as post-processing step to be applied to the CMB with the crucial aim of reducing noise and perhaps more importantly foreground contamination. It is also important to point out that, similarly to the \textit{global} Wiener filter, LIW-Filtering is - at this acronym indicates - a linear filtering technique. This means that LIW-Filtering is also relevant when studying errors and their propagation on the estimated CMB map via Monte-Carlo simulations are unavoidable.\\
Future work will focus on refining the modeling of noise and foreground residuals. Without losing the simplicity of this approach, we will more specifically focus on studying more adapted signal representation to better model the noise/foreground contamination spatial behavior.\\
 The developed IDL code will be released with the next
version of ISAP (Interactive Sparse astronomical data Analysis Packages) via the web site:\\ \\
{\centerline{\texttt{http://jstarck.free.fr/isap.html}}}

\section*{Acknowledgments}
This work was supported by the French National Agency for Research (ANR -08-EMER-009-01) and
the European Research Council grant SparseAstro (ERC-228261).

\bibliographystyle{IEEEtran}
\bibliography{Biblio_0711}

\end{document}